# CASTAway: An Asteroid Main Belt Tour and Survey.

| Author | Email | Affiliation |
|---|---|---|
| N. E. Bowles | Neil.Bowles@physics.ox.ac.uk | Department of Physics, Clarendon Laboratory, Parks Road, Oxford, OX1 3PU, UK |
| C. Snodgrass | colin.snodgrass@open.ac.uk | School of Physical Sciences, The Open University, Milton Keynes, MK7 6AA, UK |
| A Gibbings | alison.gibbings@ohb.de | OHB System AG, Universitätsallee 27-29, D-28359 Bremen, Germany |
| J. P. Sanchez | jp.sanchez@cranfield.ac.uk | Space Research Group, Centre of Autonomous and Cyber-Physical Systems, Cranfield University, MK43 0AL, UK |
| J. A. Arnold | Jessica.arnold@physics.ox.ac.uk | Department of Physics, Clarendon Laboratory, Parks Road, Oxford, OX1 3PU, UK |
| P. Eccleston | paul.eccleston@stfc.ac.uk | STFC/RAL Space, Rutherford Appleton Laboratory, Harwell Campus, Didcot, OX11 0QX, UK |
| T. Andert | tom.andert@unibw.de | Institute of Space Technology and Space Application (ISTA), Bundeswehr University Munich, Neubiberg, Germany |
| A. Probst | a.probst@unibw.de | Institute of Space Technology and Space Application (ISTA), Bundeswehr University Munich, Neubiberg, Germany |
| G. Naletto | giampiero.naletto@unipd.it | 1. Department of Physics and Astronomy, University of Padova, Via F. Marzolo 8, 35131 Padova (Italy)<br>2. CNR-IFN UOS Padova LUXOR, Via Trasea 7, 35131 Padova, Italy |
| A. C. Vandaele | A-C.Vandaele@aeronomie.be | Planetary Aeronomy, Royal Belgian Institute for Space Aeronomy, Av. Circulaire 3, B-1180 Brussels |
| J. de Leon | jmlc@iac.es | Instituto de Astrofísica de Canarias, C/ Vía Láctea, s/n, E38205 - La Laguna (Tenerife). España |
| A. Nathues | nathues@mps.mpg.de | Max Planck Institute for Solar System Research, Justus-von-Liebig-Weg 3, 37077 Goettingen, Germany |
| I. R. Thomas | ian.thomas@aeronomie.be | Planetary Aeronomy, Royal Belgian Institute for Space Aeronomy, Av. Circulaire 3, B-1180 Brussels |
| N. Thomas | nicolas.thomas@space.unibe.ch | Physikalisches Institut, University of Bern, Switzerland |
| L. Jorda | laurent.jorda@lam.fr | Laboratoire d'Astrophysique de Marseille, 38 rue Frédéric Joliot-Curie, 13388 Marseille Cedex, France |
| V. Da Deppo | vania.dadeppo@ifn.cnr.it | CNR-IFN Padova, Via Trasea 7, 35131 Padova, Italy |
| H. Haack | hhaack@mainemineralmuseum.org | Mineral and Lapidary Museum of Henderson County, Inc., 400 N. Main Street, Hendersonville, NC 28792, USA |
| S. F. Green | Simon.Green@open.ac.uk | School of Physical Sciences, The Open University, Milton Keynes, MK7 6AA, UK |
| B. Carry | Benoit.Carry@oca.eu | Université Côte d'Azur, Observatoire de la Côte d'Azur, CNRS, Laboratoire Lagrange, France |
| K. L. Donaldson Hanna | Kerri.DonaldsonHanna@physics.ox.ac.uk | Department of Physics, Clarendon Laboratory, Parks Road, Oxford, OX1 3PU, UK |
| J. Leif Jorgensen | jlj@space.dtu.dk | Technical University of Denmark, Elektrovej, Building 327, room 122, 2800 Kgs. Lyngby, DK |
| A. Kereszturi | kereszturi.akos@csfk.mta.hu | Research Centre for Astronomy and Earth Sciences, Hungary |
| F. E. DeMeo | fdemeo@mit.edu | Department of Earth, Atmospheric, and Planetary Sciences, Massachusetts Institute of Technology, Cambridge, MA, USA |
| M. R. Patel | Manish.Patel@open.ac.uk | School of Physical Sciences, The Open University, Milton Keynes, MK7 6AA, UK |
| J. K. Davies | John.Davies@stfc.ac.uk | UK Astronomy Technology Centre, Royal Observatory, Blackford Hill, Edinburgh EH9 3HJ, UK |
| F. Clarke | Fraser.Clarke@physics.ox.ac.uk | Department of Physics, Clarendon Laboratory, Parks Road, Oxford, OX1 3PU, UK |
| K. Kinch | kinch@nbi.ku.dk | Niels Bohr Institute, University of Copenhagen, Øster Voldgade 5-7, 1350 Copenhagen, Denmark |
| A. Guilbert-Lepoutre | aguilbert@obs-besancon.fr | CNRS, UTINAM - Université Bourgogne Franche Comté UMR 6213, 25000 Besancon, France |
| J. Agarwal | agarwal@mps.mpg.de | Max Planck Institute for Solar System Research, Justus-von-Liebig-Weg 3, 37077 Goettingen, Germany |
| A. S. Rivkin | Andy.Rivkin@jhuapl.edu | Johns Hopkins Applied Physics Laboratory, 11100 Johns Hopkins Rd, Laurel MD 20723, USA |
| P. Pravec | petr.pravec@asu.cas.cz | Academy of Sciences of Czech Republic, Astronomical Inst, Fricova 298, 251 65 Ondrejov, Czech Republic |
| S. Fornasier | Sonia.Fornasier@obspm.fr | LESIA, Observatoire de Paris, PSL Research University, CNRS, Univ. Paris Diderot, Sorbonne Paris Cite', UPMC Univ. Paris 06, Sorbonne Universites, 5 Place J. Janssen, 92195 Meudon Principal Cedex, France |
| M. Granvik | mgranvik@iki.fi | Department of Physics, P.O. Box 64, 00014 University of Helsinki, Finland |
| R. H. Jones | rhian.jones-2@manchester.ac.uk | School of Earth and Environmental Sciences, University of Manchester, Manchester M13 9PL, UK |
| N. Murdoch | Naomi.Murdoch@isae.fr | Institut Supérieur de l'Aéronautique et de l'Espace (ISAE-SUPAERO), Université de Toulouse, 31055 Toulouse Cedex 4, France |
| K. H. Joy | katherine.joy@manchester.ac.uk | School of Earth and Environmental Sciences, University of Manchester, Manchester M13 9PL, UK |
| E. Pascale | enzo.pascale@astro.cf.ac.uk | Cardiff University, School of Physics and Astronomy, Queen's Buildings, 5 The Parade, Cardiff, CF24 3AA, UK; Sapienza Università di Roma |
| M, Tecza | matthias.tecza@physics.ox.ac.uk | Department of Physics, Clarendon Laboratory, Parks Road, Oxford, OX1 3PU, UK |
| J. M. Barnes | Jenny.Barnes@physics.ox.ac.uk | Department of Physics, Clarendon Laboratory, Parks Road, Oxford, OX1 3PU, UK |
| J. Licandro | jlicandr@iac.es | Instituto de Astrofísica de Canarias, C/ Vía Láctea, s/n, E38205 - La Laguna (Tenerife). España |
| B. T. Greenhagen | Benjamin.Greenhagen@jhuapl.edu | Johns Hopkins Applied Physics Laboratory, 11100 Johns Hopkins Rd, Laurel MD 20723, USA |
| S. B. Calcutt | simon.calcutt@physics.ox.ac.uk | Department of Physics, Clarendon Laboratory, Parks Road, Oxford, OX1 3PU, UK |
| C. M. Marriner | Charlotte.marriner@physics.ox.ac.uk | Department of Physics, Clarendon Laboratory, Parks Road, Oxford, OX1 3PU, UK |
| T. Warren | tristram.warren@physics.ox.ac.uk | Department of Physics, Clarendon Laboratory, Parks Road, Oxford, OX1 3PU, UK |
| I. Tosh | ian.tosh@stfc.ac.uk | STFC/RAL Space, Rutherford Appleton Laboratory, Harwell Campus, Didcot, OX11 0QX, UK |

## Abstract

CASTAway is a mission concept to explore our Solar System's main asteroid belt. Asteroids and comets provide a window into the formation and evolution of our Solar System and the composition of these objects can be inferred from space-based remote sensing using spectroscopic techniques. Variations in composition across the asteroid populations provide a tracer for the dynamical evolution of the Solar System. The mission combines a long-range (point source) telescopic survey of over 10,000 objects, targeted close encounters with 10 – 20 asteroids and serendipitous searches to constrain the distribution of smaller (e.g. 10 m) size objects into a single concept. With a carefully targeted trajectory that loops through the asteroid belt, CASTAway would provide a comprehensive survey of the main belt at multiple scales. The scientific payload comprises a 50 cm diameter telescope that includes an integrated low-resolution ($R = 30 – 100$) spectrometer and visible context imager, a thermal (e.g. 6 – 16 µm) imager for use during the flybys, and modified star tracker cameras to detect small (~10 m) asteroids. The CASTAway spacecraft and payload have high levels of technology readiness and are designed to fit within the programmatic and cost caps for a European Space Agency medium class mission, whilst delivering a significant increase in knowledge of our Solar System.

## Keywords

## 1. Introduction.

Variations in composition across the asteroid population can provide a tracer for the dynamical evolution of the Solar System. CASTAway is a mission concept to explore our Solar System's Main Asteroid Belt (MAB), which can provide a comprehensive survey of the objects within the boundary conditions of e.g. an ESA Medium Class mission (ESA, 2016a). CASTAway (Comet and Asteroid Space Telescope – Away [in the main belt]) combines a long-range (point source) telescopic survey of thousands of objects, targeted close encounters with 10 – 20 asteroids (Section 3) and serendipitous searches, into a single mission concept. With a carefully targeted trajectory that loops through the MAB, CASTAway will provide a comprehensive survey of the main belt at multiple size scales. Specific science questions that CASTAway seeks to address include:

1. How do asteroid surface compositions relate to meteorite mineralogy?
2. How does the measured surface composition of asteroids vary?
3. What is the evidence for different degrees of heating?
4. How do the results from visible wavelengths "mega-surveys" (e.g. Gaia (e.g. Delbo et al., 2012), Large Synoptic Survey Telescope (LSST) (Jones, Juric and Ivezic 2015) etc.) correlate with composition?
5. How do surface composition, morphology and regolith cover vary?
6. Is our understanding of surface ages correct?

This paper describes the overall science case for a survey mission to the MAB, with necessary extrapolation to the space-based observations that will be required in the late 2020s and 2030s, a typical time scale for planning and operating a space mission. It will then discuss how practical constraints due to mission timescale, spacecraft and ultimately, available budget, lead to a compact (~800 kg) design with a targeted suite of three primary instruments that utilises an optimised trajectory to provide a comprehensive survey.

## 2.0 Scientific motivation for CASTAway

### 2.1 Small bodies & planet formation

The orbits of minor bodies (comets and asteroids) are controlled by interactions with the planets, as can be seen in the present day structure of the MAB between Mars and Jupiter. Their orbits may also trace where the planets have been in the past. When combined with compositional measurements, which largely reflect the original formation location in the Sun's protoplanetary disc, the distribution of comets and asteroids can put some of the strongest constraints on dynamical models of planet formation and evolution. Recently, numerical simulations such as the so-called Nice and Grand-tack models (Figure 1) have been successful in recreating many observed properties of the Solar System, including the distributions of relatively rocky or icy small bodies that formed at different distances from the Sun in approximately their current pattern (Walsh et al., 2011; Morbidelli et al., 2015), but there are still puzzles. Our understanding of the composition of asteroids is still very limited: Broad 'spectral types' are defined based on the shape of spectra, usually in only the visible wavelength range, but only a few thousand of the larger asteroids (from a total population of billions) have been observed. The fundamental connection between these asteroid observations and the laboratory samples we have (meteorites) is approximate and only partially understood. Understanding the composition of a broad range of asteroids with different dynamical, physical and surface properties is necessary to effectively use them to trace Solar System evolution.

The current state-of-the-art in asteroid science is detailed in the recent review book *Asteroids IV* (Michel et al., 2015). In one of the first chapters, the editors include a table that brings together a list of key open questions in current asteroid science (with references to relevant later chapters), and suggestions on how they might be addressed in the coming years (DeMeo et al., 2015). These questions form an excellent starting point for considering future avenues of exploration: For a mission in the 2030s, we expect three major topics to be important: i) **Compositional diversity in the MAB and link to meteorites**; ii) **Where is the water? What degree of heating did asteroids experience?** Iii) **Asteroid surfaces as a record of Solar System evolution; size distribution of small asteroid impactors**. The CASTAway mission is designed to address these via the following key questions, which are also summarised in Table 1 and linked to the measurement objectives in Table 2:

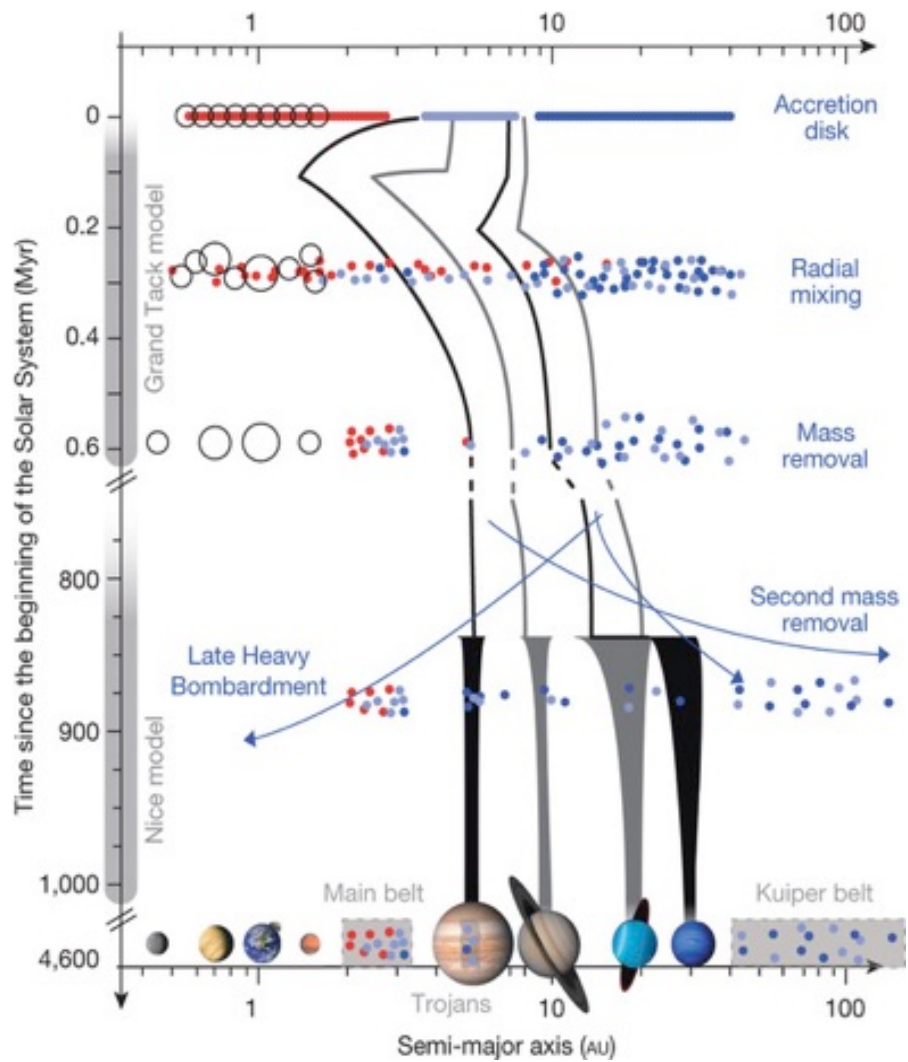

**Figure 1:** Cartoon of current models of Solar System planet formation and migration, and the effect on mixing rocky (red) and icy (blue) components in minor bodies. Reproduced from DeMeo and Carry (2014).

- In order to understand the connection between asteroids and meteorites, we need to be able to link asteroids and samples across all size ranges, from the surface samples returned by spacecraft, to samples of interplanetary dust, micrometeorites and meteorites, all the way to the characteristics of larger asteroids seen by big surveys.
    - How do measured surface compositions of asteroids vary across scales in asteroid size from metres to 100s of km; within the same taxonomic classes, families or pairs; with apparent surface age; and across different geomorphology on resolved surfaces?
    - How do surface compositions relate to meteorite types, and therefore mineralogical classification?
- In order to address the importance of the asteroid belt for questions such as the chemical composition of the Earth and the source of Earth's water, we need to understand the distribution of minerals, organics and water in the asteroid belt.
    - In the age of mega-surveys like LSST and Gaia, we will have very large numbers of broadband colours and low-resolution visible spectra of asteroids. How are these datasets best correlated with the compositional data obtained by spectra over the 0.3 – 5 μm region and the identification of water-bearing and silicate minerals?
    - How are dynamical models of Solar System formation and evolution constrained by the present day distribution of bodies containing processed minerals (heated chondritic asteroids; fragments

of differentiated bodies), organics and/or ice, i.e. what is the evidence for different degrees of heating in different regions of the early Solar System?
- In order to understand the physical and dynamical evolution of asteroids as tracers throughout Solar System history, we need to be able to interpret surface features and their relationship to location in the asteroid belt.
    - Is our understanding of surface ages, and therefore the evolution of asteroids, correct? What is the typical size distribution of craters and of small impactors, and does this agree with current theory?
    - How do surface composition, morphology and regolith cover vary with size, shape and spin rate; with the presence or absence of satellites; across different dynamical environments?

A related subsidiary question, which will be answered in addressing those above, is: How does the better-studied near-Earth object (NEO) population (and therefore the source of meteorites or samples returned by missions) sample the parent MAB, and original formation locations?

Table 1. Science questions addressed by CASTAway

| | Science question | Measurement required |
|---|---|---|
| 1 | How do surface compositions relate to meteorite mineralogy? | Measure spectra across variety of asteroid types and surface ages |
| 2 | How do measured surface compositions of asteroids vary? | |
| 3 | What is the evidence for different degrees of heating? | Map water/volatiles/organics/heated minerals distributions |
| 4 | How do visible wavelength mega-surveys correlate with composition? | Link spectra with parallel visible photometry |
| 5 | How do surface composition, morphology and regolith cover vary? | Image a variety of asteroids |
| 6 | Is our understanding of surface ages correct? (Collisional history) | Image small craters and measure size distribution of small impactors |

Table 2. Science Traceability Matrix. Question numbers refer to Table 1

| Measurement | Requirement (Threshold) | Requirement (Baseline) |
|---|---|---|
| sitivity to observe t sources to do a cant survey | Faint limit sufficient so that >10,000 objects are visible over the mission | S/N > 50 in all wavelength bins at V=15 in 1 hour | S/N > 100 in all wavelength bins at V=15 in 20 min |
| nt morphological units on asteroids | Spectral spatial resolution 10 – 20 m | 20 m | 10 m |
| cale morphological ally resolved asteroids | Imaging spatial resolution 10 – 20 m | 20 m | 10 m |
| onal maps | Narrow band imaging in flybys | S/N > 10 in 5 filters during flyby | S/N > 30 in 7 filters during flyby |
| on a variety of asteroid | Flybys of multiple asteroids | > 10 flybys over nominal mission lifetime | ~20 flybys over mission |
| Photometry and -board software | Detection of nearby asteroids to V=16 | Detection of a statistically significant sample: ~1000 per size bin over the mission | Orbit determination for discovered asteroids; follow-up spectroscopy |
| ensity through trajectory | Doppler tracking of the spacecraft throughout a flyby | Flyby distance of ~1000 km. Power loss due to HGA mis-pointing < 0.1 dB | Flyby distance < 100 km |
| tral features due to OH. | Photometry at 0.3 µm | Detect water production rate Q > $10^{26}$ molecules s$^{-1}$ (Ceres) in < 10 min at S/N>3 | Detect water production rate Q > $10^{23}$ molecules s$^{-1}$ (MBC) in < 1 hour at S/N>3 |
| metry at visible | Photometry in *ugriz* filters | S/N > 20 in 1 min | S/N > 50 in 1 min |
| ure maps | Surface temperature mapping between 150 – 300 K | Accuracy of 5 K, spatial resolution of 20 m | Accuracy of 1 K, spatial resolution of 10 m |
| aps in the thermal | Photometry in thermal-IR filters | S/N > 10 in 14 filters during flyby | S/N > 30 in 14 filters during flyby |

## 2.2 The composition of asteroids

Asteroid compositions can be inferred through spectral measurements, albedos, and densities. Following earlier photometric work, dedicated spectroscopic surveys were started in the 1980s. Focusing first on the visible part of the spectrum, with low spectral resolution, these surveys led to a new definition of asteroid spectral classification in the early 2000s (Bus et al., 2002). This was followed by surveys in the near infrared (NIR) up to 2.4 μm, allowing further refinements of the classification by DeMeo et al. (2009), which is the current reference (Figure 2). Asteroid spectral taxonomies in the visible and NIR generally separate asteroids into a few major complexes, each made up of a number of classes. The C and the S complexes make up the majority of bodies in the MAB, while a host of smaller classes capture some of the variety of compositions, such as the spectrally very red D-types, or the basaltic V-types that make up the Vesta family.

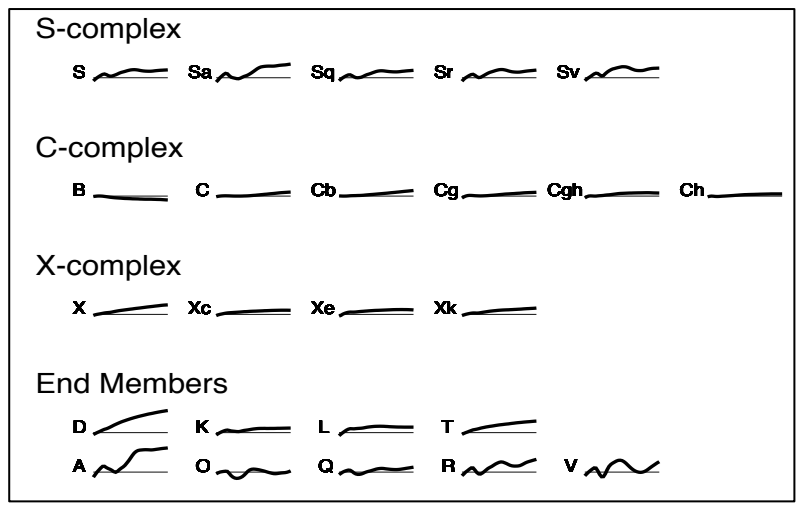

Figure 2. Bus-DeMeo taxonomy (DeMeo et al. 2009). Shape of reflectance spectra over 0.4-2.4 μm for various classes.

By classifying large numbers of asteroids through spectral or photometric surveys (Bus and Binzel, 2002; Ivezic et al., 2001) the distribution of asteroid types across the main belt can be determined. The classical example (Gradie and Tedesco, 1982) shows a clear trend in the largest asteroids from classes S, C, P, and then D as a function of increasing distance from the Sun, which was interpreted as a relic of the initial temperature gradient at the time of formation. Yet, after almost 30 years of spectroscopic surveys of asteroids from the Earth, only about 4500 asteroids have been characterised, and most only in the visible. Several groups have focused on the 3 μm and the 8 to 40 μm regions, and suggest different sub-classes with different compositions (Takir and Emery, 2012; Rivkin et al., 2012), but these samples remain limited to a few tens or hundreds respectively (Emery et al., 2006), due to the limits of current telescopes and/or the Earth's atmosphere.
Our understanding of the distribution of material in the asteroid belt, from families to overarching structure, was improved by mining various broadband sky surveys. These suggest more mixing of asteroid types at smaller sizes than the simple variation with heliocentric distance seen earlier (Parker et al., 2008; DeMeo and Carry, 2014). Nevertheless, while such surveys improve the sample size, they do not improve our understanding of asteroid composition, which they inherit from spectroscopic surveys.

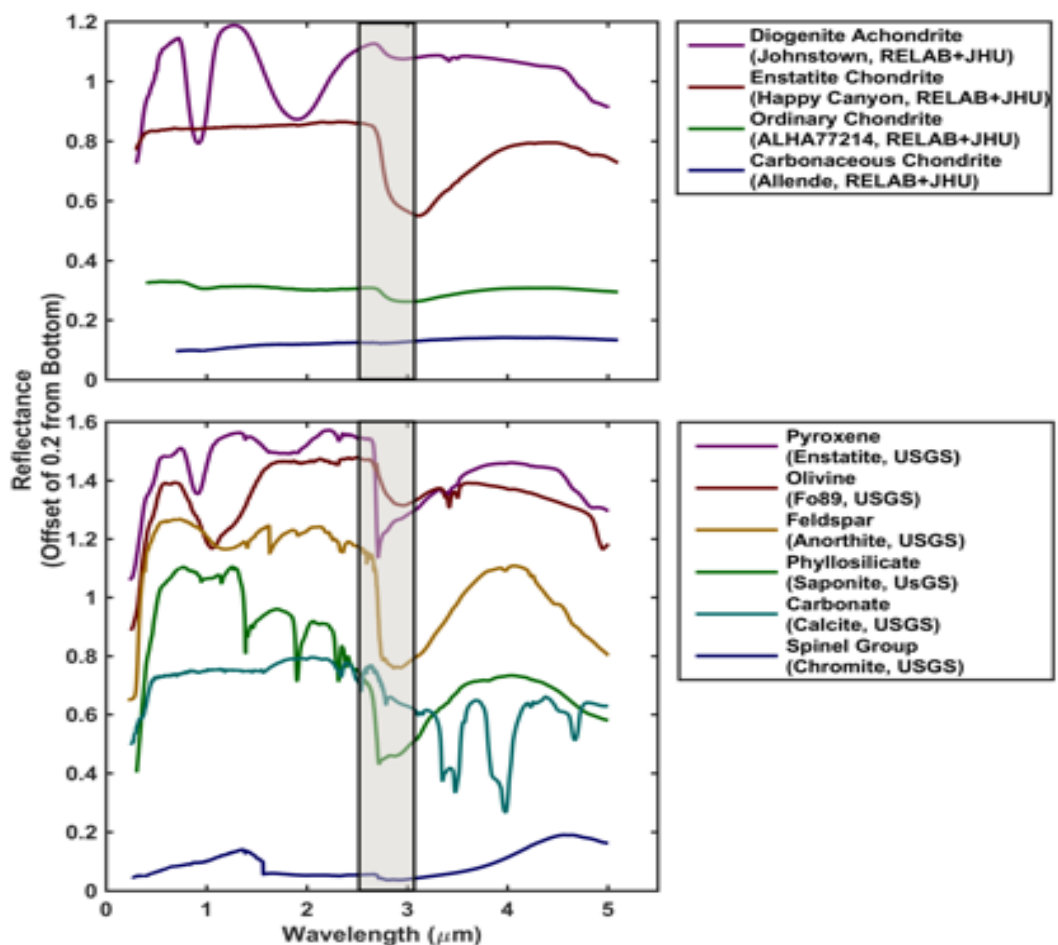

Figure 3: spectra of various meteorites and minerals over 0.3 – 5 µm. Grey box indicates atmospheric absorption region.

Absorption features in visible and NIR spectra due to various minerals are typically broad (Figure 3, Table 3) Features related to water and hydrated minerals are found around 0.7 and 3 µm, but do not always correlate with each other (Rivkin et al., 2015). Water ice itself also has features at 1.4 and 2 µm, but these have been identified on (1) Ceres only, in observations from the Dawn spacecraft (Combe et al., 2016, Nathues et al., 2017). There are also a variety of absorption features around 3 µm due to organic molecules. The presence of organics (DeSanctis et al., 2017), water and/or hydrated minerals is an important diagnostic for asteroid formation and history, as it implies that material must have been incorporated from parts of the proto-planetary disc that were distant from the Sun. Recent observations point towards a significant amount of water in the MAB; at Ceres (Russell et al., 2016), Themis (Rivkin and Emery, 2010; Campins et al., 2010), and (implied) in the main belt comets (MBCs) (Hsieh and Jewitt, 2006; Jewitt et al., 2015).

Table 3. Main spectral features of common minerals.

| Minerals | Main Spectral Features (µm) |
| --- | --- |
| **silicates** | |
| olivine (forsterite, fayalite) | 0.86-0.92, 1.05-1.07, 1.23-1.29 |
| pyroxene (enstatite, pigeonite) | Mg-Fe: 0.91-0.94, 1.14-1.23, 1.80-2.07; Ca-Mg-Fe: ~ 1, 1.2, 2 |
| feldspar (anorthite = Ca endmember) | Na-Ca: 1.1-1.29 |
| phyllosilicate (e.g. saponite) | ~ 1.35, 1.8, 2.3, 2.8 |
| phyllosilicate (e.g. serpentine) | ~ 1.4, 1.9, 2.2, 2.9 |
| phyllosilicate | ~0.7, 0.9, 1.1 |
| **oxides** | |
| chromite | 0.49, 0.59, 1.3, 2 |
| spinel | 0.46, 0.93, 2.8 |
| **carbonates** | |
| calcite / dolomite | 2.50-2.55, 2.30-2.35, 2.12-2.16, 1.97-2.00, 1.85-1.87, ~ 3.4, ~ 4 |
| **phosphates** | |
| apatite | OH-apatite: 1.4, 1.9, 2.8, 3; F-Cl apatite 2.8, 3.47, 4.0, 4.2 |
| **organics** | |
| e.g.(n-alkanes, amino acids) | numerous e.g. ~1.7, ~2.3, ~2.4 (for alkanes) |

Much more detailed compositional information for asteroids can be measured, via various techniques, on laboratory samples. Asteroid samples include the small amount of material from sample return missions and the large collection of meteorites, the majority of which originate from asteroids. Asteroidal meteorites are divided into distinct groups based on their mineralogy and chemical compositions (Krot et al., 2014). The iron, stony-iron and achondrite meteorites originate from bodies that were hot enough for melting and differentiation to take place. However, most meteorites (>90%) are chondrites derived from asteroids that were smaller and/or formed slightly later, resulting in less extensive heating. Some chondrites preserve pristine materials from the birth of the Solar System, whereas many others were affected by thermal metamorphism and/or aqueous alteration on the asteroidal parent bodies.

A comprehensive understanding of the meteorite record requires detailed knowledge about the asteroids from which they originated. Similarly, direct comparisons with meteorite spectra are necessary in order to interpret accurately the surface mineralogy of asteroids. In some cases the links between asteroids and meteorites are robust. The most abundant meteorite group, the ordinary chondrites, is linked to a subset of the common S-type asteroids, whilst the Howardite-Eucrite-Diogenite (HED) meteorites are related to the V-type asteroids. In contrast, spectra of dark C-type asteroids have fewer and weaker characteristic features, making it challenging to infer their compositions. The C-type asteroids are usually linked to the carbonaceous chondrites, but resolving the exact nature of this relationship remains a key goal in planetary science (Reddy et al., 2015). Detailed matching of asteroids with the wide range of individual meteorite groups is limited: Without this link we are unable to connect meteorite mineralogy with source regions.

## 2.3 Current and future observations

The greatest limitation on spectroscopic observation of asteroids from Earth is the terrestrial atmosphere, especially when trying to search for water or hydration related features, with the 2.5 – 2.9 µm gap in most spectra falling at one of the most interesting regions in the NIR, and ozone absorbing most flux at the 0.3 µm OH emission band. At 3 µm, only ~100 km scale asteroids are bright enough to observe, and most have already been measured (Rivkin and Emery, 2010).
Surveys such as the ESA Gaia and Euclid missions, and the upcoming LSST, are expected to increase the sample by an order of magnitude (Delbo et al., 2012; Carry et al., 2016; Jones et al., 2015), but only in broad-band photometry and/or at visible wavelengths. They will not provide the NIR spectroscopy necessary to improve our understanding of asteroid composition. The James Webb Space Telescope will have a spatial resolution of ~75 km in the MAB, and will be able to obtain NIR spectroscopy to 5 µm on all currently numbered asteroids (with a limit of V~23) (Rivkin et al., 2016). NIR spectroscopy with the E-ELT at V~24 (sub-km asteroids in the MAB) would take 20 – 30 minutes per object, although much of the crucial 3 µm spectral region will still be blocked by Earth's atmosphere (Figure 4). There will be very limited time available for any one topic with these large multi-purpose telescopes, so observing a significant fraction of the vast population of small asteroids that will be discovered by LSST and Gaia will not be possible.

The very smallest asteroids (1 – 10 m diameter) will remain beyond the detection limit of LSST, and would be challenging for ELTs even if we knew where to look. However, these bodies are most likely the direct parents of the meteorites we collect on Earth. Spectroscopy of these bodies (probably single boulders without a coating of regolith) would be very valuable in linking astronomical observations, and therefore all the larger surveys, with laboratory measurements.

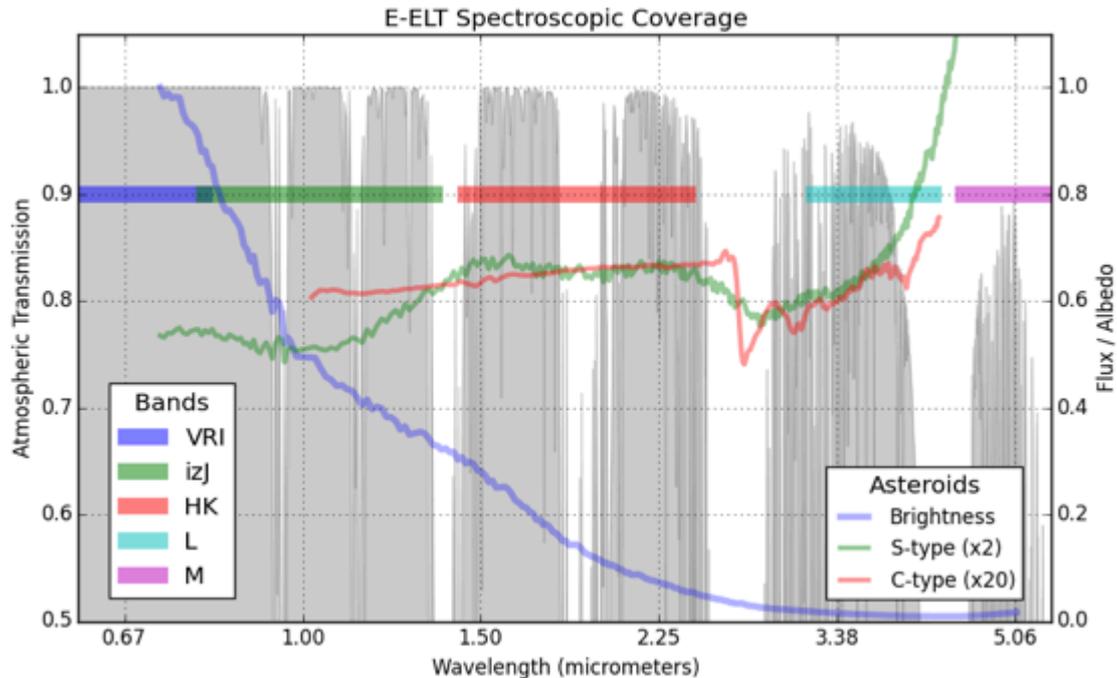

Figure 4: Earth atmospheric transmission (grey) and spectroscopy bands planned for the EELT, compared with asteroid spectra. The observed spectrum (blue) is similar in all cases on this scale - dividing by the solar spectrum to show the reflectance spectrum reveals the differences between an S-type (green) and C-type (red). Data from missions to Gaspra (Granahan 2011) and Ceres (De Sanctis et al. 2015).

Finally, except for the largest main belt asteroids (e.g. Ceres (Li et al., 2011)) surveys with any ground-based or near-Earth telescopes will not resolve asteroids at the scale where they can detect compositional or morphological variation across the surface. To do this there is no alternative but to visit the asteroids with spacecraft.

To date spacecraft have visited 12 asteroids **(Table 4)**, but in only four cases has the asteroid been the primary mission objective, with the spacecraft entering orbit. However, flybys at least partly optimised for asteroid science showed the potential of such encounters.

**Table 4. Asteroids visited by spacecraft (D = diameter, CA = Spacecraft Closest Approach, classes after the Bus-DeMeo taxonomy (DeMeo et al., 2009).**

| Name | D (km) | Class | Explored by | Date | CA (km) |
|---|---|---|---|---|---|
| 1 Ceres | 952 | C | Dawn – Orbiter | 2015- | 200 |
| 4 Vesta | 529 | V | Dawn – Orbiter | 2011-2012 | 200 |
| 21 Lutetia | 120 x100 x80 | Xc | Rosetta – Flyby | 2010 | 3162 |
| 243 Ida | 56×24×21 | Sw | Galileo – Flyby | 1993 | 2390 |
| 253 Mathilde | 66×48×46 | Cb | NEAR Shoemaker – Flyby | 1997 | 1212 |
| 433 Eros | 34×11×11 | Sw | NEAR Shoemaker – Orbiter / lander | 1998-2001 | 0 |
| 951 Gaspra | 18×10.5×9 | S | Galileo – Flyby | 1991 | 1600 |
| 2867 Šteins | 4.6 | Xe | Rosetta – Flyby | 2008 | 800 |
| 4179 Toutatis | 4.5×~2 | Sq | Chang'e 2 – Flyby | 2012 | 3.2 |
| 5535 Annefrank | 4.0 | S | Stardust – Flyby | 2002 | 3079 |
| 9969 Braille | 2.2×0.6 | Q | Deep Space 1 – Flyby | 1999 | 26 |
| 25143 Itokawa | 0.5×0.3×0.2 | Sq | Hayabusa – Orbiter / Sample Return | 2005 | 0 |
| 162173 Ryugu | ~1 | Cg | Hayabusa 2 – Orbiter / Sample Return | 2018 | 0 |
| 101955 Bennu | ~0.5 | B | OSIRIS-REx – Orbiter / Sample Return | 2018 | 0 |

The asteroids visited to date demonstrate a wide variety of surface structures (e.g., Murdoch et al., 2015). They are heavily cratered bodies with a coating of fine-grained loose regolith. The majority of smaller asteroids examined to date show evidence (from morphology, shape and density measurements) of 'rubble-pile' structure (Fujiwara et al., 2006), although there is a population of asteroids with (partially) differentiated interiors and higher densities (Sierks et al., 2011). In spatially resolved spectral data there is evidence for compositional variation at various scales (Russell et al., 2016; Belton et al., 1996 and papers therein).

Craters on asteroids are important tracers of their history and interior structure (material strength), with crater counting being the only means to date asteroid (or planetary) surfaces. To calibrate these surface age measurements it is necessary to know the population of impactors – i.e. very small asteroids, down to metres in diameter in the case of craters 10s of metres in size. In the MAB these very small bodies are not detectable from Earth, so age estimates in this region (or even on the surface of Mars) are based on the measured population of small NEOs, and assumptions on how this can be compared with the MAB. A direct measurement of the size distribution of the smallest asteroids would therefore calibrate the chronology of planetary surfaces, a fundamental measurement in our understanding of Solar System evolution. As discussed above, the very smallest asteroids (probably single boulders) are also of great interest as a direct link between meteorites and larger bodies, because they should be free of any regolith layer. All visited asteroids, even sub-kilometre ones, have been found to have a significant regolith layer; larger asteroids seem to have a finer, or deeper regolith than small ones (Delbo et al., 2007).

## 2.4  Evolution of asteroids

The present day population of asteroids is the result of 4.5 billion years of collisional and dynamical evolution (Morbidelli et al., 2015). Only the very largest asteroids appear to have escaped catastrophic (disruptive) impacts; the many asteroid families observed in the MAB are fragments from major collisions (Nesvorny et al., 2015). Current theories suggest that the original planetesimals in the MAB were relatively large bodies, and that the asteroids we see today are the outcome of a cascade of collisional fragmentation (Morbidelli et al., 2009), together with reforming of fragments into rubble pile asteroids. Bodies that were larger than ~100 km in diameter, and which formed soon after the first Solar System solids, probably experienced internal heating from radioactive decay of short-lived radioisotopes in the first tens of millions of years after accretion. The extent of heating was variable, but was sufficient in some cases to at least partially melt and differentiate the body, meaning that fragments from different depths within these original bodies have different compositions (including metallic iron from the core). Testing how composition varies throughout collisional families, or between pairs of asteroids identified as having recently split apart (Margot et al. 2015), can test how much variation existed within single parent bodies. Pairs and young families are particularly interesting, as their surfaces will better represent the interior composition of their parent bodies. Also, young surfaces are important because they have not been exposed to space weathering for as long.

## 3.  Science Requirements for a Main Asteroid Belt Survey Mission

Many of the important open areas in asteroid science can be addressed with a series of relatively simple measurements. The key requirement is to significantly increase the sample of asteroid compositional and morphological data, covering the size range from D~1 m to D~100 km. This will link laboratory samples (from meteorites and NEO surface sample return missions in the 2020s) with large (primordial) asteroids. To do this requires a large-scale spectroscopic survey that covers the critical wavelength ranges for mineralogy (visible – 5 μm), measuring the size distribution in the MAB down to meteorite parent sizes (1 – 10 m diameter), and obtaining morphology and composition information at comparable resolution for a varied sample of asteroids. These objectives are described in more detail below. They can all be achieved with one efficient mission by placing a small space telescope in the MAB, where it will combine a large spectroscopic survey of asteroids with multiple close flybys, and a programme to discover very small asteroids.

The first requirement in improving our understanding of the MAB is to significantly increase the number and variety of asteroids visited by spacecraft. Resolved imaging and spectroscopy reveal both the structure of the asteroid, any compositional variation across the surface, and also allow a unique comparison of composition with surface age, via crater counting and comparing spectra from different geomorphological areas (e.g. within relatively fresh craters vs. old surface terrain). Encounters by spacecraft are also the only way to make mass measurements for small, non-binary, asteroids, necessary to test internal structure models by comparing the bulk density with the grain density of minerals identified by surface spectroscopy (Consolmagno et al., 2008; Scheeres et al., 2015). The flybys should be relatively close, which leads to design of the spacecraft and instruments around flybys with ~1000 km closest approach distance. At this distance one kilometre on the surface corresponds to 0.06°, so a typical flyby target will fit comfortably within a 1° FOV (17.5 km). To enable counting of the small craters to study the surface age, and to resolve compositional differences across bigger craters (e.g. walls vs. floor vs. surrounding area), requires a spatial element on the surface of ~10 m, corresponding to ~2 arcsec/pixel. A narrow angle camera with a 2k x 2k detector would achieve both of these requirements for the geomorphology goals.

For the composition goals, wavelength coverage across the visible and NIR is necessary. Coverage from 0.3 – 5 μm would allow any UV drop off to be seen, the visible spectral slope to be measured, characteristic absorption features from 0.7 to 3.5 μm to be detected, and the beginning of the thermal IR emission to be characterised between 4 and 5 μm. The thermal properties are so important to understanding the geophysics of the surface layer that a longer wavelength channel, covering e.g. 6 – 16 μm, is essential for interpreting the resolved surface data. This wavelength range also contains further absorption features that are complementary to the NIR range, and is widely used in the study of meteorites. This implies the need for three detectors, covering the UV/visible, NIR and thermal IR respectively. As the NIR region is most diagnostic for composition, spectroscopy is the preferred technique in this range, with a slit spectrometer operating in push-broom mode being the obvious approach for mapping the surface as the spacecraft moves across it (nadir tracking). At shorter and longer wavelengths imaging is more appropriate, firstly to enable straightforward interpretation for geomorphology goals, but also as a selection of narrowband filters would be sufficient to capture the available compositional information.

To connect the detailed information returned by flybys to the asteroid population as a whole, and to construct a detailed compositional map of the MAB, it is necessary to obtain spectroscopic measurements of a significant fraction of the population, covering all dynamical classes and sizes. A survey of at least 10 000 asteroids covering the visible and NIR would be enough to map composition throughout the MAB in a statistically meaningful way, covering ~1% of all known asteroids, increasing our current knowledge by orders of magnitude. This number will provide significant numbers of objects in sub-groups of different dynamical (orbital zones, asteroid families), compositional (spectral types) and physical groups (e.g. size bands, activity). In addition, the range of sizes of asteroids for which spectra will be obtained will be extended to much smaller values than is possible from Earth-based measurements. As the visible range contains relatively few absorption features and it is the spectral slope that is primarily of interest, low spectral resolution is sufficient, for example using broadband photometry in the SDSS filters, to match the LSST survey band-passes. A spectral resolution in the NIR of $R$~100 is sufficient to identify broad absorption features, and, crucially, to distinguish between different minerals and ices with features around 3 μm.

This survey should include all larger asteroids and a representative selection of smaller ones, with the capability to obtain spectra over this wavelength range for sub-km asteroids. To do this from Earth orbit would require a large space telescope (JWST class), but by placing the telescope in the MAB small asteroids can be targeted from ~0.1 AU range at visual magnitude V<15, requiring only a modest telescope with 50 cm diameter. Being within the MAB also offers the opportunity to get spectra on very small asteroids (down to metre-sized) via target-of-opportunity reaction to new discoveries, as these will also regularly pass close enough to appear with V<15.

One of the key questions to answer in mapping the composition of the MAB is "where is the water?" In addition to the NIR absorption features due to surface ice and/or hydrated minerals, it is also important to look for evidence of subsurface ice via comet-like activity (escaping water vapour or dust lifted by this gas). The NIR spectral range includes emission bands from water around 2.7 μm, and there is a very strong feature from OH (a direct photodissociation product of water) at 0.3 μm. The latter can be isolated via narrowband filters in a blue-sensitive CCD imager, with the advantage that detecting this via imaging would allow integration over the full size of any potential gas coma (many arcminutes when observed from ~0.1 AU range).

It is also worth noting that at this typical distance, assuming an imager with the 2 arcsecond/pixel scale required for the flyby mode above, the spatial scale of 1 pixel will be 150 km, which is almost comparable to the HST (50 km for a main belt asteroid at perigee) and far superior to seeing-limited ground-based observations (1000 km). This is very valuable for studying the dust escaping from MBCs/active asteroids, as the high spatial resolution is crucial to (1) detect small fragments embedded in the debris trail or orbiting the main nucleus, (2) measure typical sub-escape speed dust velocities from the out-of-plane extent of the dust tail, and (3) detect very low activity levels from Point Spread Function (PSF) broadening. A series of active asteroid observations with the HST (Jewitt et al., 2010; Jewitt et al., 2014; Agarwal et al., 2013; Agarwal et al., 2016) have shown that such high resolution provides constraints on the dynamics of the ejected material that is not accessible from ground, while the unique observing geometries possible with observations from within the MAB can also be useful in studying these dynamics, even when observing potential MBCs from > 0.1 AU (Snodgrass et al., 2010).

Although only ~20 active asteroids/MBCs are currently known, the rate of discovery has increased with the Pan-STARRS survey and is expected to increase again with LSST. The most recent estimates suggest that there are 50-150 objects with activity somewhere in their orbit at a similar brightness to the currently known MBCs (Hsieh et al., 2015), and it is reasonable to assume at least an order of magnitude larger population at fainter levels that will be detectable with LSST. While close encounters will still be rare, opportunities to observe weakly active bodies with at least comparable resolution to large ground-based telescopes should occur regularly throughout the mission. Independent of this, having a large sample of asteroids imaged at good S/N with a stable (out-of-atmosphere) PSF will allow a careful and consistent search for activity in all survey targets, and therefore stronger constraints on the total population of MBCs.

The final piece of the puzzle in understanding asteroid evolution and linking them with meteorite samples is discovering 1 – 10 m diameter asteroids. This is the population that impacts larger asteroids to produce the 10s to 100s of metre diameter craters seen in flybys and rendezvous missions, and measuring the size distribution is therefore critical for measuring surface ages. Also, these are the direct parents of meteorites and are likely to be regolith-free, making them important targets for spectroscopy. These will not be detectable from Earth with even the largest telescopes, but should be very common (~$10^{14}$ and $10^{12}$ 1 and 10 m diameter bodies are expected, respectively) and therefore pass relatively close to a spacecraft in the MAB on a regular basis, on average there being one within ~8000 km or ~60 000 km for 1 – 10 m size bins. Discovering them requires a camera with sufficient sensitivity in exposures of a few seconds, and on-board software capable of recognising sources and comparing them with known stars and asteroids: Advanced star-tracker type cameras can provide both of these (Section 6.5). There is a trade of shallow and wide field or deeper but narrower pencil beams for relatively distant objects, with the latter approach being more efficient. This also has the advantage that, by aligning the search window with the main spectroscopy survey telescope, it will be possible to respond to discoveries with a small (<1°) slew. Such a search plus target-of-opportunity mode is the only way to obtain spectra of very small asteroids.

An additional aspect to studying the populations of smaller asteroids is that it also allows us to build the connections between Near Earth Asteroids (NEAs), the source for meteorites, and their original parent populations (e.g. Binzel et al., 2015). Although meteorites, as small objects in the same population, are believed to share common sources with NEAs from a number of dynamical resonances in the MAB, the asteroids in this population are likely to have come to these regions within the last few million years. The Yarkovsky effect allows small asteroids to drift into resonances that are distant from the orbits of their original parent bodies over timescales of tens to greater than hundreds of My, as shown by the cosmic ray exposure ages of meteorites. Thus the original source regions of the NEAs could be anywhere in the asteroid belt (e.g. Bottke et al., 2002). The relatively constant flux of the NEA population is further evidence for "trickle" feeding of resonances from distributed sources. Thus a survey across the MAB will enable an important comparison with the NEA population and the recent meteorite record. A comprehensive spectroscopic survey and 1 – 10 m census will therefore provide a unique way of determining if the NEAs and meteorites are representative of the population as a whole.

## 4.0   Overview of Trajectory Design Options.

The design of a suitable trajectory for CASTAway plays a key role in achieving all the previously discussed science objectives. This section summarises the process of searching and identifying potential trajectories for CASTAway.

### 4.1. Trajectory requirements and boundary conditions.

CASTAway's trajectory requirements and constraints are derived from two different sources; the mission science objectives and ESA's boundary conditions for the 5th call of medium size missions (M5, ESA 2016a) within the Cosmic Visions programme. The following sections describe how these two sources of top-level requirements and constraints lead us to a preliminary design of CASTAway's trajectory.

The discovery of very small (1 – 10 m) asteroids requires the spacecraft to be within ~100 000 km, for a 10 m class object. CASTAway shall thus orbit within the MAB in order to find a statistically significant number. The asteroid detectors (Section 6.5) are expected to discover on the order of one small asteroid a day, on average. Using cumulative Poisson distribution to model the significance of the discovered sample, and defining as *significant* a 99% confidence that the number of counts should be within 10% of the average-based estimate, nearly 1000 days of small asteroid survey are required to reach the 99% confidence-level.

The Main Telescope (i.e., for the spectroscopic survey, Section 6.2) will need to be within the MAB in order to survey <10 km class objects. Assuming ~1 hour per target for spectroscopy and allowing about 20% time for overheads, calibration, etc., detailed compositional information of a minimum of 10 000 objects could be attained in about 500 days of survey operations[1]. This would include ~10% of all currently known main belt asteroids with a radius >10 km. Together, the survey and small asteroid discovery operations place a requirement for the minimum operational time in the MAB to be approximately 3 years.

The morphological and geological study of some pre-selected set of ≥10 asteroids requires spatially resolved spectral mapping and visible light imaging. This unequivocally requires CASTAway to encounter asteroids at close-range, i.e., ~1000 km. Obtaining mass estimates via range-rate measurements and orbital perturbation also requires the spacecraft to encounter the asteroid at a similar distance range.

CASTAway was initially proposed to ESA's 5th call for Medium-size missions ("M5"), and thus, it was designed to satisfy all the specific boundary conditions of the call. Particularly, ESA's Cost at Completion (CaC) for medium-size mission was capped to 550 M€ (2016). This implies that, under typical ESA cost breakdown, the launch cost must be on the order of 70 M€, and consequently European medium-lift launchers were considered.

At the time of writing, the current European medium-lift launcher is Soyuz. However, the Ariane 6.2 (A62) is scheduled to replace Soyuz by ~2024. Unfortunately, at this time there remains uncertainty on the actual launch capabilities of A62. Based on an extrapolation from the A62's Geostationary Transfer Orbit (GTO) capability, as reported in the draft Ariane 62 User Manual (ESA, 2016b), the escape energy (C3) performance will be somewhat better than that of Soyuz. However, the Soyuz's performance is adopted as a worst-case scenario for the CASTAway trajectory design. If feasible mission scenarios that meet CASTAway's science goals (Section 2, 3) can be found for a Soyuz-like performance, then significantly better options ought to be available for an A62 launch that outperforms Soyuz.

---

[1] The 10 000 figure is based on our current knowledge of the MAB. If by launch the targets are available then up to 44 000 objects could be surveyed, including observation time for flybys.

In summary, the CASTAway trajectory design assumes a direct insertion into an Earth escape trajectory by an A62 with Soyuz-like performance, hereafter referred to simply as Soyuz launch. The values of mass inserted into the hyperbolic escape trajectory as a function of $v_\infty$ (i.e. C3) and declination $\delta_\infty$ used are those reported in the M5 technical annex (ESA, 2016a). A trajectory for CASTAway shall then be sought that spends >3 years within the MAB and encounter ≥10 MAB objects at a distance in the order of 1000 km. Moreover, the visited objects should include a large variety of compositional types, sizes and orbital regions. In order to limit operational costs, the total CASTAway mission duration was limited to 7 years but could be extended.

**4.2. Preliminary trajectory analysis.**

This section provides a simple analysis based solely on fundamental astrodynamics principles. This analysis allows us to gain context to the above description of requirements and constraints, as well as some initial insights into what constitutes the feasible trajectory design space for CASTAway. Figure 5 shows the potential spacecraft mass that could be delivered into an orbit with periapsis at 1 AU and apoapsis distance $Q$. Figure 5 is computed assuming a variable hyperbolic escape velocity $v_\infty$ in the Earth's velocity direction. 'Nominal' performance refers to the performance that can be achieved with the reported launcher capability (ESA 2016b, ESA 2016, Perez, 2012), while 'Extended' assumes a single deep space manoeuvre (DSM), soon after departing Earth, using a bipropellant engine with specific impulse of 320 s. Figure 5 also provides insight to the potentially achievable benefits if A62 capability outperforms Soyuz by (e.g.) up to 50% in terms of payload mass inserted into an escape trajectory, i.e. improved C3 capability. This scenario is thereafter referred to as A62(S+). Low thrust propulsion was discarded on the grounds of complexity and cost (Section 5.3). A launch vehicle adapter of ~75 kg was assumed (Gibbings et al., 2016; Perez 2011). Finally, the performance is only extended to that allowed by a minimum spacecraft dry mass of 800 kg.

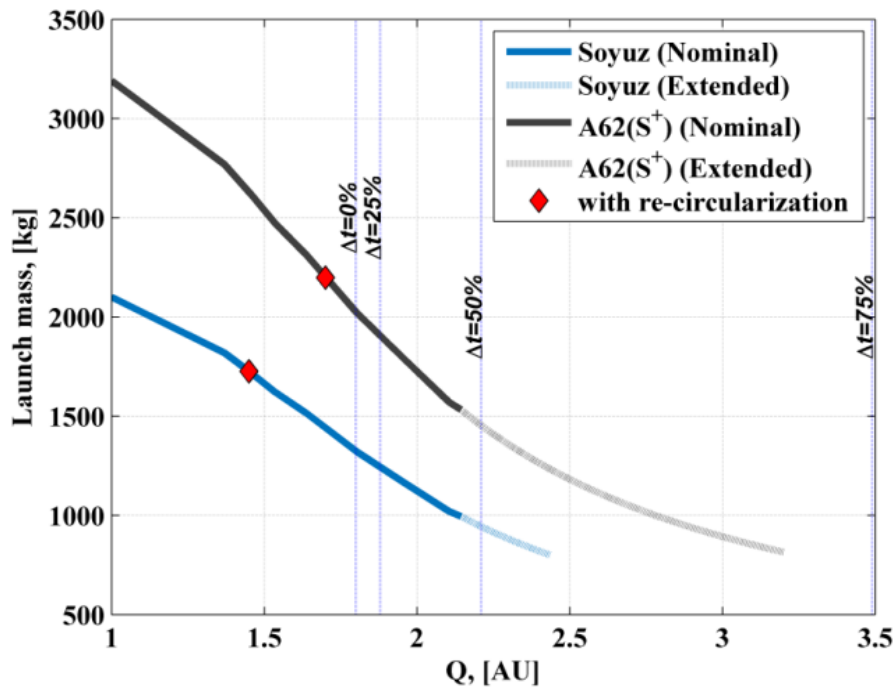

Figure 5: Launch mass for Soyuz and A62(S+) as a function of apoapsis $Q$. Continuous lines indicate reachable apoapsis with a direct insertion, while dashed lines indicate the extension possible with a DSM soon after departing Earth. Curves extend to 800 kg dry mass. Four vertical lines indicate apoapsis for which a total of 0%, 25%, 50% and 75% of the time is spent within the asteroid main belt.

Firstly, and most importantly, note that a Soyuz launch can deliver nearly 1 tonne of payload mass into an orbit with an apoapsis at 2.21 AU allowing the spacecraft to spend 3.5 of a 7 years mission beyond 1.8 AU. Such a mission scenario should likely meet the science objectives for compositional and small asteroid surveys, while also complying with the M5 boundary conditions. However, there is a significant difference in Figure 5 between the reachable apoapsis for Soyuz and A62(S+) cases. This clearly indicates that any increase of performance, even marginal, from current Soyuz capability would result in a larger science return or enhancement, in terms of capability to explore deeper into the MAB.

Figure 5 also shows the maximum orbital radius that can be reached, considering a circularization burn at apoapsis, i.e. a circular-to-circular Hohmann transfer. It can be noted that for both the Soyuz and the A62(S+) performance this orbital radius lies well below the MAB. This however does not consider the possibility of using a Multi Gravity Assist (MGA) trajectory to reach a circular orbit within the MAB. A global search of MGA trajectories was also performed (Cano, 2016; Sánchez et al., 2016). The search demonstrated that, due to the relatively short mission duration (7 years), direct insertion into interplanetary orbits with relatively high C3 (~20 km$^2$/s$^2$) was the optimal choice. One single gravity assist (GA), either of Mars or the Earth, increases the operational time spent within the MAB by about 20%, with respect to that of sequences without GAs.

In summary, the optimal option for CASTAway's trajectory was identified as a low eccentricity heliocentric orbit. However, the analysis also points out the limited Δ$v$ budget that can be allocated for manoeuvring within the MAB. Nevertheless, as described in the following section, or in more detail by Sánchez et al. (2016), hundreds of 10-asteroid fly-by sequences were found compliant with all the requirements described above.

### 4.3. Global Trajectory Search.

The design of the CASTAway trajectory presents an extremely challenging multi-objective mixed-integer programming problem. Similar problems have been proposed as challenges to be solved during the Global Trajectory Optimization Competitions (GTOCs)[2]. In particular, the CASTAway trajectory shall maximise the quality of the surveys (i.e. number of new detections and spectral data), as well as the number of asteroid fly-bys. The trajectory must fly by 10 or more objects, and the sample of visited asteroids must include a wide range of asteroid types and sizes.

It is clear that the design of an adequate operational orbit for CASTAway presents some unique challenges; most particularly, solving the large discrete combinatorial problem that is required to find feasible asteroid fly-by sequences. While some available literature exists on methods to tackle these problems (e.g.Grigoriev and Zapletin, 2013), a new methodology was designed specifically for the CASTAway mission study. This methodology is described in detail in Sánchez et al., (2016) and allowed the exploration of the entire design space for two different trajectory types: Firstly, trajectories considering a Soyuz launch and encountering 10 MAB objects with no planetary gravity assists (GA) and, secondly, trajectories considering also a Soyuz launch and 10 asteroid fly-bys, but adding one single Mars GA.

The outcome of this global search is a catalogue of 232 different MAB tours, each with 10 asteroid flybys. All of these trajectories were found to be feasible within the M5 boundary constraints, assuming a launch performance of a Soyuz, and were found within a reduced asteroid database of ~100 000 MAB objects, which contained an adequate diversity of asteroids in size and orbital distribution. However, note that the number of known asteroids is expected to increase by an order of magnitude by the time of the launch of the M5 mission due to significant new surveys such as the LSST (Jones et al., 2015) and Gaia (e.g. Delbo et al., 2012).

These 232 MAB sequences explore 1348 different asteroids, thus some repetition occurs in the catalogue. Figure 6 shows how these 1348 objects cover the different regions of semi-major axis – eccentricity space. The encounters belonging to trajectories using a Mars GA are marked with red squares, while black dots identify sequences without GA.

---

[2] http://sophia.estec.esa.int/gtoc_portal/

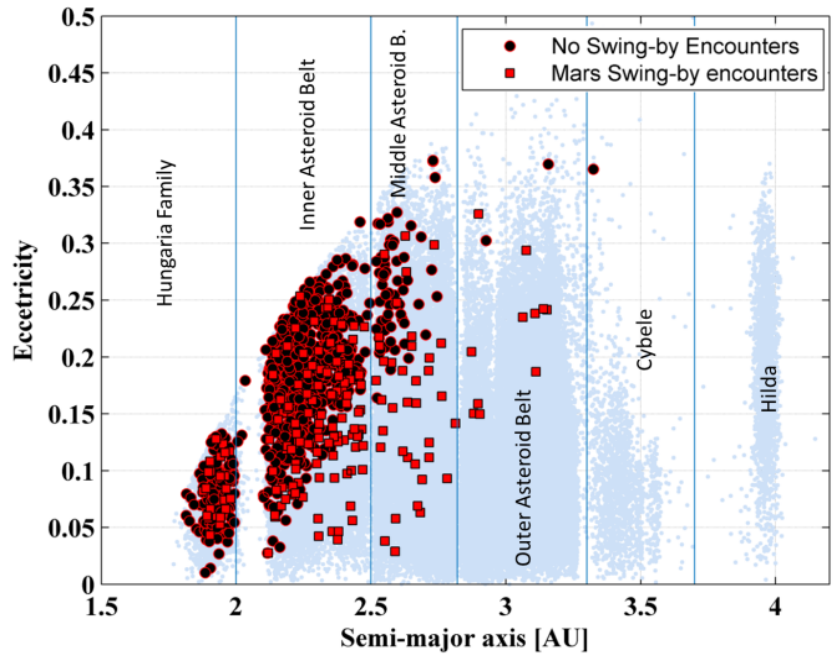

Figure 6: Semi-major axis and eccentricity map of the complete catalogue of encounters. Small background dots represent the available objects in the search database (~100,000). Each of the 232 sequences meets 10 asteroids.

The set of solutions found allows a good spread of asteroids within the Hungaria family, the inner and middle main belt, including objects belonging to the Flora, Vesta, Nysa, Maria and Eunomia families. The number of objects visited in the outer main belt is limited, although a small increase in launcher performance would easily enable a good spread of targets also in the outer belt (see section 5.3).

## 5. Space Segment

The scientific objectives (Section 3) of the CASTAway mission can be achieved within a relatively simple space segment. The analysis has been confirmed with a study, using the concurrent engineering design facilities at OHB system AG in Bremen, Germany. Conservative assumptions were applied throughout, which was fully compliant to ESA's margin philosophy for science assessment studies (ESA, 2014). The following text outlines the design that resulted from this first iteration. It is based on technology from previous ESA missions (BepiColombo, ExoMars and Rosetta) and well-characterised concepts from Echo (M3), ARIEL (M4), Lisa Pathfinder and the James Webb Space Telescope.

### 5.1. Mission Analysis

The baseline trajectory was identified among the 232 solutions described in Section 4.3. This particular trajectory was selected because of the interest and variety of the asteroids visited, the total time spent within the MAB and also to ensure specific engineering criteria such as the timespan between each asteroid fly-by or the avoidance of critical mission phases during superior solar conjunctions. The summary of the baseline trajectory is represented in an inertial heliocentric reference frame in Figure 7.

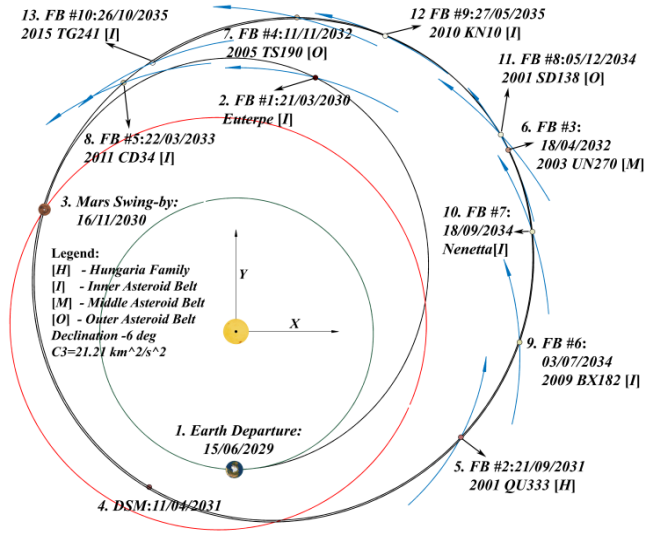

Figure 7: CASTAway baseline trajectory for the OHB Spacecraft design.

The launch vehicle, an Ariane 6.2 with assumed Soyuz-like performance, inserts the spacecraft into a hyperbolic Earth escape trajectory with $v_\infty$ of 4.60 km s$^{-1}$ and declination of –6°. According to ESA (2016a Figure 1) this represents the insertion of nearly ~1250 kg into an escape trajectory, from which nearly 175 kg would be wet mass for the CASTAway system (i.e. ~75 kg of launch adapter). After commissioning of the spacecraft into an Earth escape trajectory, CASTAway will nominally require 11 more deep space manoeuvres (DSMs) to complete the tour (orbital correction manoeuvres not included). The first, and largest of these would be performed two weeks after departing the Earth's gravity well, requiring nearly 500 m s$^{-1}$ of $\Delta v$. This manoeuvre could potentially be reduced substantially, subject to the performance of the launcher over the declination ($\delta_\infty$) range of [–6$^o$, –10$^o$]. The manoeuvre also includes a 3$\sigma$ injection correction $\Delta v$ of 57 m s$^{-1}$, to account for launcher insertion errors. The 10 remaining DSMs are substantially smaller, ranging from 120 to 10 m s$^{-1}$, and are planned after each celestial body encounter (i.e. Mars and asteroids). The total $\Delta v$ performed by the on-board propulsion system is of 990 m s$^{-1}$, including 100 m s$^{-1}$ for AOCS (100% margin) and 5% margin for all manoeuvres computed by trajectory optimization in the patched-conic dynamical framework (ESA, 2014).

Figure 8 summarizes the timeline of an entire 7-year trajectory. The first flyby event occurs before performing the Mars gravity assist, visiting (27) Euterpe (an S-type asteroid). The remaining nine flybys occur while the spacecraft completes two further loops of the heliocentric orbit. Thus, CASTAway completes a total of three crossings of the MAB. In the first crossing, CASTAway will reach a distance of 2 AU from the Sun, while the two subsequent crossings, after the Mars swing-by, the spacecraft reaches an apogee of about 2.5 AU.

The Mars swing-by also substantially increases the total mission time spent within the MAB. The Mars minimum altitude during planetary swing-by was allowed to be 250 km altitude, taking into account ESA mission operation heritage, specifically that of Rosetta's successful Mars gravity assists at 250 km altitude on the 25$^{th}$ February 2007 (Ferri et al., 2008). Mars GA increases both periapsis, which is set to 1.4 AU, and apoapsis of CASTAway's operational orbit, which consequently allows access to a larger set of asteroid families and regions in the MAB.

Fly-by operations, which include optical navigation and implementation of trajectory correction manoeuvres (TCM), would start as soon as the asteroid is clearly visible on the Visible Context Imager (Section 6.2.2) in navigation mode. This should allow for the planning and execution of TCM manoeuvres several weeks before each fly-by. Thus, fly-bys should be planned so that sufficient time is allocated between them, as well as approach conditions that enable the detection of the asteroid several weeks in advance of the fly-by. For the baseline trajectory, for example, FB#6, FB#7 and FB#8 occur 77 days apart from each other, requiring manoeuvres of 27 m s$^{-1}$ and 11 m s$^{-1}$ to correct and aim for FB#7 and FB#8 respectively. The mentioned 5% margins allocated for each DSM is compliant with ESA margin philosophy for science assessment studies (ESA, 2014). It should also be noted that TCMs for the Rosetta fly-bys of (2867) Steins and (27) Lutetia were both well below 1 m s$^{-1}$ (Accomazzo et al., 2010, 2012) as a guide to the margin requirement.

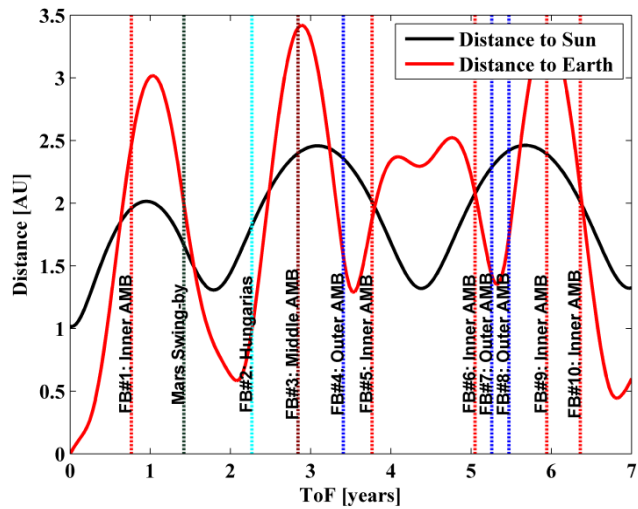
Figure 8: Baseline trajectory timeline and distance to Earth and Sun.

### 5.2. Spacecraft System

Figure 9 shows the spacecraft in its deployed and launch configuration respectively. The spacecraft has a total wet mass at launch, including all margins, of about 1150 kg. The payload contribution is about 80 kg, together with about 340 kg of propellant and a 75 kg launch adapter. A launch mass margin of 4 % on top of the generous sub-system margins makes the design robust towards any potential mass growth and uncertainty in the launch performance. Additional trade-off criteria are discussed in (Gibbings et al., 2016).

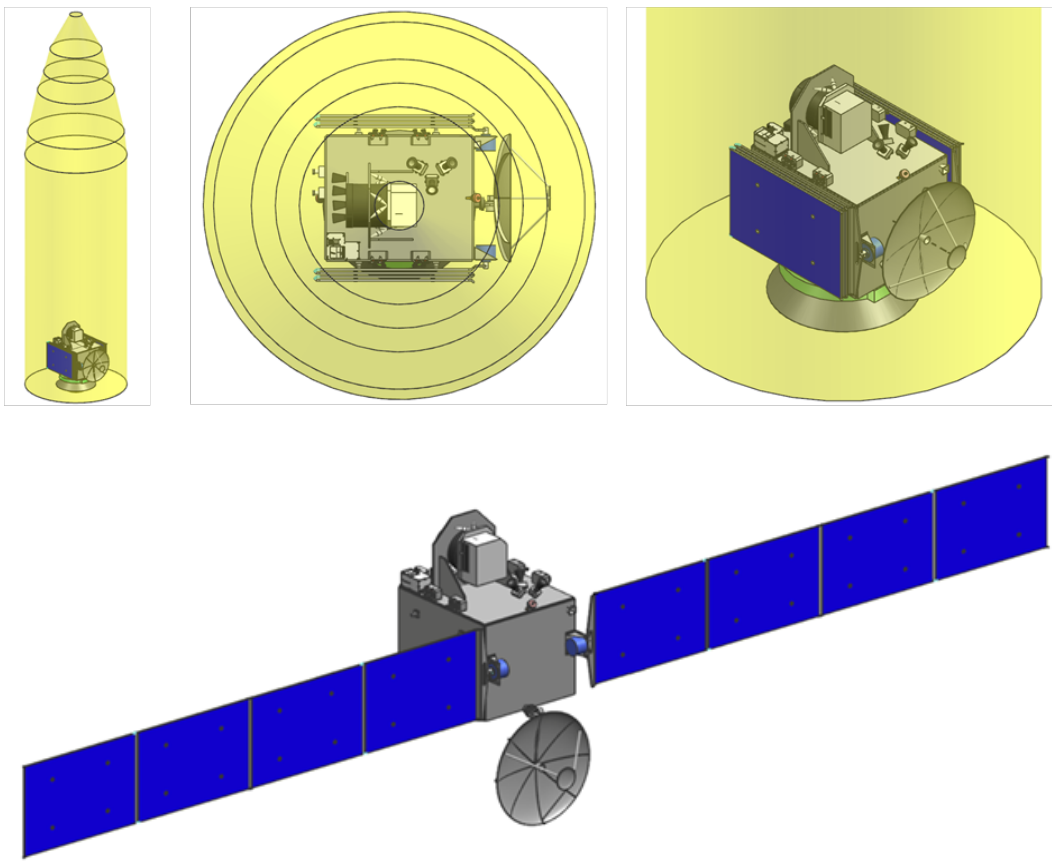

Figure 9: Spacecraft - Launched and Deployed Configuration (image credit: OHB System AG)

The payload pallet is accommodated on the top deck of the spacecraft. All three instruments have the same boresight orientation. The launcher interface ring is located on the base. The deployable solar arrays are accommodated symmetrically on opposite sides of the spacecraft. A steerable High Gain Antenna (HGA) is accommodated on the panel facing in the opposite direction to the instrument boresight. The solar array wings can be rotated for Sun tracking (1 Degree Of Freedom (DOF)). The HGA is equipped with a two axis pointing mechanism to enable independent Earth pointing capabilities.

The spacecraft´s structure consists of an aluminium alloy honeycomb with aluminium facesheets. A boxed shear-web structure is connected to the launcher interface ring, and establishes the primary load-bearing path. Additional sandwich panels create compartments for internal accommodation. The propulsion subsystem follows a standard MON (Mixed Oxides of Nitrogen)/MMH (monomethylhydrazine) bi-propellant chemical approach. Two sets of small-Newton thrusters provide full 3-axis attitude control capability. Another set of two thrusters execute all the mission manoeuvres (i.e. flybys and swing-bys). Two propellant tanks (MON & MMH) are accommodated in the central shear web. Helium is used as a pressurant for the bi-propellant and is stored in a smaller tank. The power subsystem includes two conventional solar arrays. Each are sized for when the spacecraft is in survey mode, when the payload and communication system are operated simultaneously within the MAB. Maximum power pointing tracking provides optimised power generation. The battery is sized for launch and early operations. No other eclipses will be encountered throughout the mission. The battery is only required to power the heaters and avionics until the solar array is fully deployed and operational. The Attitude Orbital Control System (AOCS) and Guidance Navigation and Control (GNC) subsystem includes a standard suite of star trackers, Sun sensors and reaction wheels. These units are combined with a dedicated navigation camera, an inertial measurement unit, and a high performance gyroscope. Attitude anomaly detectors also provide payload protection from critical (Sun) illumination conditions. Pointing and long-term stability, rather than agility, drives the AOCS design. In survey mode the spacecraft will stare constantly at certain selected sections of the MAB, before moving on. Precise tracking is needed to follow the point-source targets. Stability is achieved with the high performance gyroscope and fine guidance signals (images) from the telescope. Optical navigation, with autonomous tracking of the asteroid's centre of brightness, is used in the flyby mode (in the same way Rosetta asteroid flybys were performed).

The communication subsystem consists of an X-band system and a steerable HGA. Two additional low gain antennas are used during launch and early operations. A wide lobe medium gain antenna is used for contingency and to establish an Earth-link in safe mode. Payload data storage is only required during the flyby events. Data is stored on the spacecraft's Data Storage and Handling Assembly (DHSA) and is downloaded over the following weeks. Data downlink occurs via the 35 m ESTRACK ground stations of New Norica, Cebreros or Malargue (ESA's tracking station network). The mass memory is capable of covering four days of science and house-keeping data. CASTAway's baseline trajectory design was designed such that no critical mission operation, such as a fly-by, occurs during solar phase angles within 177 to 183º. When the solar phase angle is near 180 degrees, the noise in the communication link increases significantly, as radio signals travel through regions of high charge-particle density near the Sun. In reality, even within less than 1 degree of Sun-Earth-Probe (SEP) communication can be maintained, although, in order to keep a favourable signal to noise ratio, data rates need to lowered. Similarly, orbital determination is also affected by solar conjunction, and this is the reason why no critical operation could be planned with 3º of SEP. From the baseline trajectory, CASTAway will enter 4 periods of solar conjunction (i.e. <3º SEP) each lasting between 6 and 17 days. Nevertheless, it is likely that CASTAway will be able to maintain sufficient data rate to keep a house-keeping data link during these stages of the mission.

A simple passive thermal control subsystem is used with a heater in a closed loop system. No louvres, heat switches or other complicated elements are required. Instead, external and internal MLI, thermal finishes, thermal doubler, sensors and interface fillers are combined with a radiator, heater and heat pipes.

### 5.3. Alternative architectures

Given the uncertainty of the Ariane 6.2 launcher performance, the feasibility of two alternative mission configurations was also assessed. The first scenario addressed the necessary increase in the launcher's uplift capability to reach an orbit far deeper into the MAB (approximately 3.2 AU with a mission Δv of 1.3 km s$^{-1}$). This has the advantage of improving the flexibility of the launch window and accessing a more scientifically compelling range of flyby targets. Several tens of trajectories with 14 asteroids were found by exploring less than 2 % of the potential design space. The second configuration addressed potential mass and cost saving measures, which could be implemented for the baseline Soyuz-like solution. It would increase the overall launcher performance margin and safeguard the mission against a reduced, worse-than-Soyuz launcher performance. For both cases the scientific payload remained the same. Only spacecraft subsystem level changes were made. An electric propulsion (EP) transfer was also investigated. It was however rejected due to its longer transfer times (reduced science operations) and higher cost. EP would also require larger solar arrays and radiators, resulting in an increased structure, thermal and propulsion subsystem. It is not compatible with the boundary conditions of the M5 mission call. Nevertheless, if an EP system was shown to be a viable solution it would enable other compelling mission profiles to be considered. Preliminary design of low thrust trajectories for MAB tours have previously demonstrated that the higher specific impulses of EP systems outweigh the disadvantages of thrust in the order of 100 mN (Di Carlo and Vasile, 2016; Tora, 2016).

A launcher uplift performance increase of approximately 23 % relative to the Soyuz baseline would be required to access an apogee of 3.2 AU (see Figure 10). The spacecraft's dry and wet mass at launch (including adapter) increases to about 850 kg and about 1480 kg respectively. Recommendations given in the M5 mission briefing by ESA suggested that such an improvement is a reasonable expectation: The performance capability to GTO and for L2 transfer orbits, for example, increases for an Ariane 6.2 by 54% and 62 % respectively when compared to the Soyuz performance.

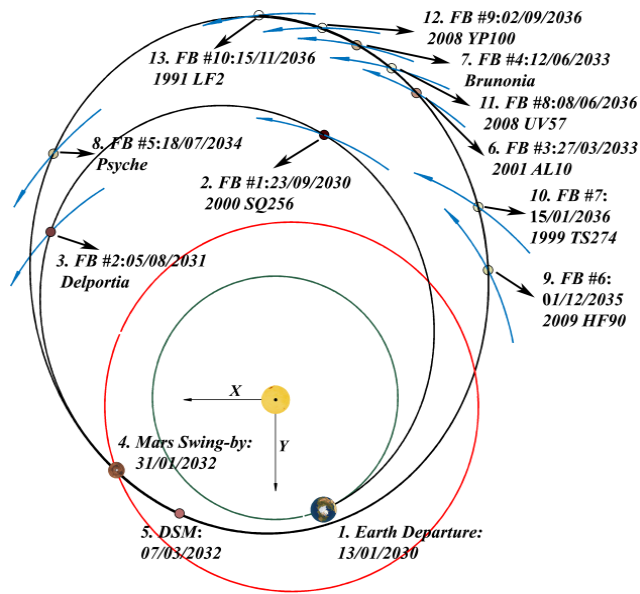

**Figure 10. CASTAway trajectory for increased performance scenario/system, visiting a large variety of regions in the MAB, as well as (in this example) the metallic asteroid (16)Psyche.**

If an increase in mission Δ𝑣 is available, giving access to greater heliocentric distances, this will impact on the propulsion and structure subsystem. The propellant mass increases to about 550 kg, which requires two slightly larger propellant tanks and a larger helium tank. The increase in tank mass increases the surrounding structure (no change in the tanks' location) and the spacecraft's height. The biggest change however occurs in the power subsystem. A 50 % increase in the solar array area is needed to account for the increase in the Sun-spacecraft distance with an appropriate decrease in the solar flux. The heater power demand and radiator area also increases. The thermal control subsystem is marginally affected by the increase of the spacecraft´s dimensions. The AOCS/GNC subsystem requires larger reaction wheels to provide a slightly higher slew rate. Additional propellant is also needed for reaction wheel offloading. There are no changes to the communication or data handling subsystems. The only compromise is the increased time (reduced timeliness) to return data to Earth due to the decreased data rate. The total launch mass increase w.r.t the Soyuz baseline is about 320 kg.

This additional scenario (increased launcher performance) demonstrated the flexibility of the CASTAway mission concept. Optional mass-saving measures are available if the performance of the new Ariane 6.2 launcher is lower than the reference Soyuz-like performance. Solutions include: 1) Using a miniature X-band transponder and low gain antenna; 2) Reducing the power amplificaition of the communication subsystem; 3) Reducing the propellant volume margin by using smaller tanks, 4) Replacing the thrusters with a 400 N liquid apogee engine for conducting the mission manoeuvers; 5) Using more efficient solar cells; 6) Removing a propulsion Remote Terminal Unit; and 7) Using a digital sensor bus to reduce the harness mass. Removing the pointing mechanism of the HGA was also considered, but rejected. It would have saved at least 15 kg (direct mass), but the spacecraft would lose the ability to perform the radio science experiment activities. Dedicated communication modes would also be needed.  Implementing these seven solutions would save about 50 kg of the spacecraft´s dry mass w.r.t the Soyuz-like baseline. The total dry mass, with system margin, is then reduced to about 680 kg.  The largest saving comes from the propellant mass. The liquid apogee engine provides a higher specific impulse (320 seconds compared to 300 seconds), saving about 60 kg. Only about 280 kg of propellant is required. The total wet mass at launch including the adapter is therefore reduced to about 1040 kg, saving about 115 kg with respect to the nominal mission platform design. The launcher mass margin is increased to about 160 kg or by about 14 %. These optional measures make the design even more robust towards a potential mass growth and uncertainty in the launch performance.

The baseline trajectory (Section 4) represents a worse case analysis for the case of Soyuz-like performance; it is expected that the A62 performance may allow an apoapsis reaching beyond 3 AU (e.g. Figure 10).  For the baseline trajectory, a necessary trade is an apospsis of 2.5 AU therefore only giving access to outer-belt asteroids with an eccentricity larger than 0.15 AU (after Sanchez et al., 2016).

## 6. Outline Science Payload
### 6.1 Payload overview

To achieve the science objectives (Section 3) the CASTAway payload includes three primary instruments (Figure 11):

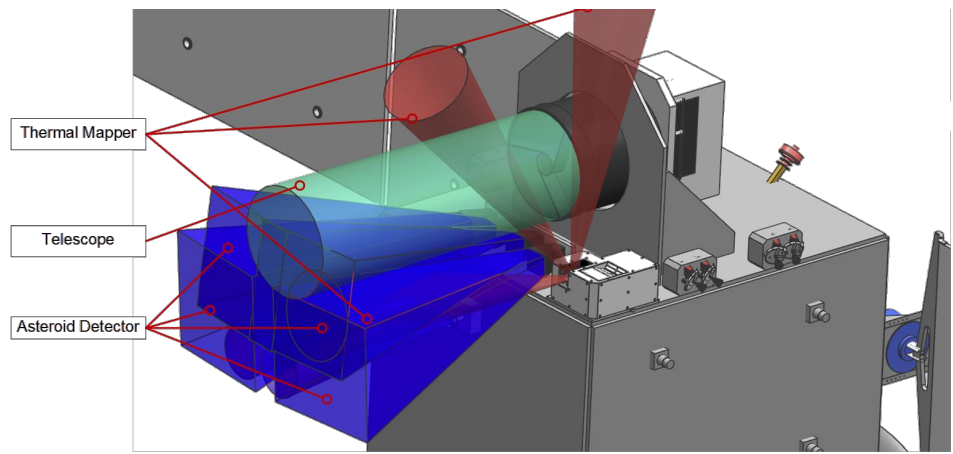

Figure 11: Payload accommodation on spacecraft with FOVs shown

A Main telescope/spectrometer for point source survey and flyby, thermal-IR (6 – 16 µm) mapper for close approach flybys and four asteroid detecting cameras derived from star-trackers.  The payload elements are mounted on a single face of the spacecraft to simplify assembly, integration and verification (AIV).  The asteroid detector cameras are aligned to have an overlapping field of view with the main instrument to provide assistance with target acquisition and tracking during survey observations, and to enable follow up observations on asteroid discoveries.

### 6.2 Main Telescope for CASTAway
#### 6.2.1 Point source and imaging spectrometer.

The primary goal of the Main Telescope for CASTAway (Figure 12,  Figure 13)  is to provide near infrared (0.6 – 5 µm), moderate resolution ($R = 30 – 100$) spectroscopy for a range of targets from distant (up to 1 AU) point sources to 10 – 20 m spatial resolution observations during flyby closest approach.

The baseline instrument is based on a series of high heritage designs (Bagnasco, et al., 2007; Neefs et al., 2015). The telescope is a 500 mm aperture modified Korsch off-axis type design with heritage from the RALCAM series of Earth observation instruments that have been through extensive ground test and qualification campaigns (Tyc et al., 2008). The light from the telescope is then dispersed into a moderate-resolution spectrum ($R = 30 – 100$) using a prism (either sapphire or MgO based) and imaged onto a Mercury Cadmium Telluride (MCT) detector array. Possible near-IR detectors include one based on the H2RG 5 µm detector used by the NIRSpec instrument on JWST (e.g. Bagnasco et al., 2007).

The H2RG detector requires cooling to ~40 K for its most sensitive operation, especially at wavelengths > 4 µm. Even at distances of > 2 AU this will require the inclusion of an active cooler to achieve this operating temperature range. Some cooling will also be required (e.g. to ~ 100 K) for the spectrometer module to reduce the background thermal load on the detector.

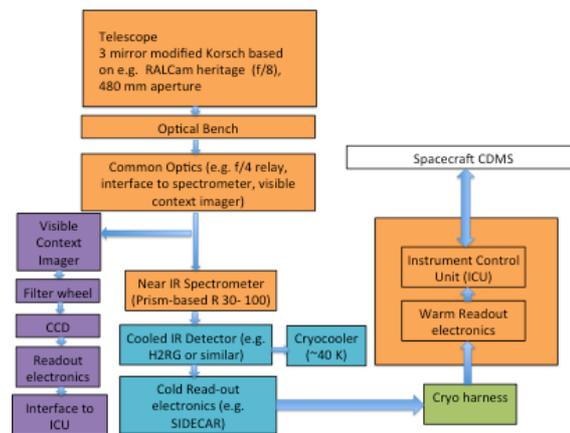

Figure 12: Main Telescope for CASTAway optical and system block diagrams

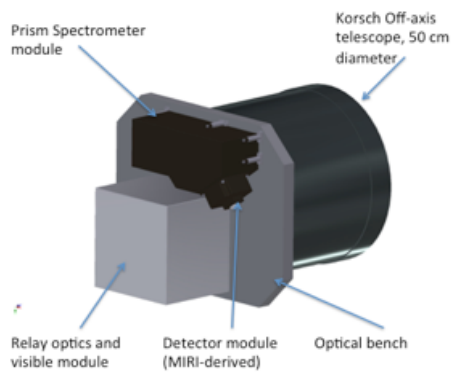

Figure 13: CASTAway Main Telescope Layout

The Main Telescope for CASTAway includes a long 1˚ slit across the focal plane to provide redundancy and astrophysical background monitoring during the point source survey and increased mapping capability during the flyby/encounter phases. Spatially resolved spectral cubes (i.e. images where each spatial element also contains spectral information) will be measured during flybys to provide compositional information on each of the target asteroids.

### 6.2.1 Main Telescope Operating Modes

#### 6.2.1.1 Science cruise point source spectroscopic survey – "Survey Mode"

During the science cruise phase between flybys the main telescope is slewed onto a point source target at the start of an observation sequence. The integration time for the near-IR detector array will be set depending on the expected flux, but given that the majority of targets are expected to be at or near $V_{mag}$ ~15 it is likely that multiple tens of seconds integrations will be required during e.g. a twenty minute observing slot. The detector would be read out in a window around the spectrum to reduce data volume.

The majority of the time that the main telescope is in Survey Mode the spacecraft will be executing a pre-determined, optimised, observing sequence i.e. a command table load with schedule. The instrument switches to "flyby" mode once the flyby target asteroid becomes spatially resolved in the spectrometer slit.

#### 6.2.1.2 Asteroid Flyby mode

In flyby mode the readout rate from the detector is steadily increased as the flux increases towards the asteroid. The spacecraft flyby motion and tracking is used to scan the target, push-broom style to map the surface of the asteroid.

During the flyby the narrow angle Visible Context Imager (Section 6.2.2) integrated into the main telescope will provide moderate spatial (~10 – 20 m) views of the surface, complementing the coverage from the slit spectrograph. The VCI images will allow re-projection of the spectrometer to match spectral and surface features (Coradini et al., 2011).

### 6.2.2 Visible Context Imager (VCI)

The visible context imager serves three main purposes; a) Resolved imaging in flyby mode, b) Target acquisition for survey spectroscopy and flybys, c) Photometry in survey mode.

It is fed from the main telescope using a dichroic beam splitter, and is a straightforward CCD imager with filter wheel(s) to provide spectral selection.

#### 6.2.2.1 Technical specifications

The VCI CCD, filter wheel and shutter are very standard parts with heritage from many missions. The important feature is that the shutter must be capable of a range of exposure times from 10s of milliseconds in flyby mode (to avoid smearing) to 10s of seconds in survey mode, but this has been achieved in many previous cameras (e.g. Rosetta/OSIRIS, Dawn/FC). Two sequential filter wheels (each containing an empty slot for use with filters in the other wheel) will probably be best, rather than one larger one. The context camera will share a common DPU with the spectrograph.

#### 6.2.2.1 Design parameters

Table 5: Filters for the VCI CCD context imager

| Filter | Central Wavelength (nm) | Width (nm) | Note |
|---|---|---|---|
| *u* | 355 | 63 | Survey mode. SDSS-type filter. |
| *g* | 469 | 141 | Survey mode. SDSS-type filter. |
| *r* | 617 | 139 | Survey mode. SDSS-type filter. |
| *i* | 748 | 154 | Survey mode. SDSS-type filter. |
| *z* | 893 | 141 | Survey mode. SDSS-type filter. |
| OH | 308 | 10 | OH emission band |
| UVcont | 300 | 10 | Continuum for OH search / UV flyby filter |
| Blue | 400 | 30 | Flyby filter (approx. spec)/Spectral Slope |
| Red | 600 | 30 | Flyby filter (approx. spec)/Spectral Slope |
| Hydra | 701 | 22 | Flyby filter. Based on Rosetta/OSIRIS NAC filter |
| NIR | 850 | 40 | Flyby filter (approx. spec) |
| $Fe_2O_3$ | 932 | 35 | Flyby filter. Based on Rosetta/OSIRIS NAC filter |

The VCI FOV and pixel scale are set by flyby considerations. To achieve the necessary spatial resolution for identifying small impact craters etc. (Section 3), and also have most of a typical sized asteroid in the FOV, it should have a 1˚ FOV and 2 arcsec pixel scale, i.e. a 2k x 2k CCD chip. These correspond to 17.5 km and 10 m respectively at the typical closest approach distance of 1000 km.

The CCD camera will also provide spectral information at wavelengths shorter than 0.6 μm, where there are no strong absorption features, but the slope of the reflectance spectrum varies between asteroid types, some showing a 'UV drop off' (DeMeo et al., 2009). At these wavelengths broadband photometry is sufficient to measure this slope, and to allow comparison with other asteroids. In flyby mode, where fluxes will be relatively high, a set of narrow-band filters will limit the throughput and also complement the spectrograph composition maps by identifying any major compositional variation that can be seen in the visible (i.e. revealed by colour changes across the surface).

#### 6.2.2.2  Filter selection

In order to allow direct comparison with the LSST survey (and other large sky surveys), the Sloan Digital Sky Survey (SDSS) *ugriz* filters will be included as part of the VCI set. Absolute calibration against field stars will also be possible using these filters. Narrowband filters for flyby mode should include those centred on the 0.7 and 1 μm absorption bands attributed to phyllosilicates and olivine/pyroxene respectively, and additional continuum filters spaced evenly across the visible range. Finally, in order to detect outgassing water from selected targets (Ceres and other large water-bearing asteroids, comets, suspected MBCs) narrow-band filters around the 308 nm OH emission band and nearby continuum are included. The continuum UV filter will also be useful in flyby mode to give the UV slope.

### 6.3  Main Telescope Performance Budgets

#### 6.3.2  Telescope temperature.

For the faintest targets (e.g. V ≈ 15) the thermal background from the telescope may become appreciable for wavelengths >3 μm, requiring background monitoring and subtraction. Operating the telescope with a minimum temperature of < 120 K will be advantageous. An initial assessment of the operating temperature of the telescope at 2.2 AU shows that this is achievable with standard thermal designs.

Initial estimates of the Main Telescope temperature using a simplified radiative balance model show that the telescope can be passively cooled to ~113 K.

More complex models with ESATAN (Figure 14) show that with shadowing from the Solar arrays and shielding from the spacecraft are taken into account, temperatures of ~100 K are possible. The pointing implications would need to be considered during mission operations planning.

It is expected that the spectrometer module will include a cooled "inner sanctum" of ~100 K to reduce the thermal background on the MCT IR detector.

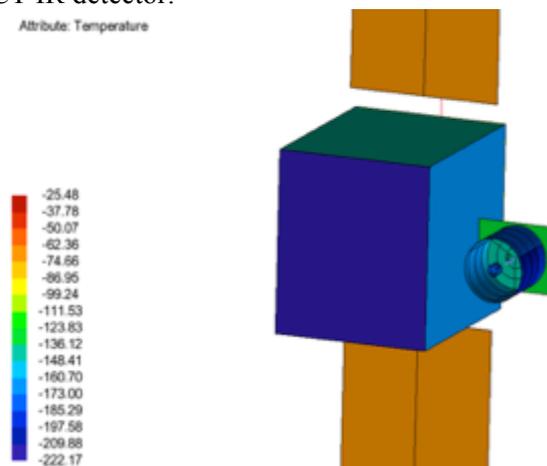

Figure 14: CASTAway Main Telescope ESATAN temperature estimated. Assumed distance is 2.5AU with the Sun incident on the Solar arrays to the right of the image.

### 6.3.3 Active Coolers for the CASTAway Main Telescope

The cooling requirements for the H2RG detectors on the CASTAway spectrometer are 80 mW at 40 K. Passive cooling at 200 K is possible in the current spacecraft design to support this. These requirements can be met using either a small scale Stirling cycle cooler or a Neon Joule Thompson cooler derived from the Planck 4 K cooler design.

### 6.3.4 Main Telescope Performance simulations

During flyby the flux across both the visible and infrared spectral ranges will increase as the spacecraft nears closest approach. The integration time of the detectors will be stepped to prevent saturation.
In the point source survey mode between encounters the telescope will be operated in a photon-counting mode with integration times set by the expected flux from the target. Given that the bulk of the targets are expected to be V ~15, integration times of order minutes to hours with on-board co-adding are expected to be required. Simulations based on experience gained from the EChO exoplanet transit spectroscopy mission Phase A study (e.g. Pascale, et al., 2015) and adapted for CASTAway have shown that spectrometer designs that maintain the necessary level of photometric stability are possible. For typical asteroid point source survey targets (V mag >=15) CASTAway is photon noise limited by the target at wavelength < 3 micron. At longer wavelengths, instrument and target photon noise contribute to the noise budget in roughly equal portions. Brighter targets are photon noise limited across the whole band.

### 6.3.4 Main Telescope Summary

Table 6. Instrument Summary, Main Telescope for CASTAway.

| Parameter | Units | Value/Description | Remarks |
|---|---|---|---|
| Reference P/L | N/A | Main Telescope | CASTAway Main Telescope for point source survey and flyby mapping |
| Spectral Range | μm | 0.6 – 3.5 (goal=5)/ 0.3 – 1.0 | Spectrometer/context imager. |
| Spectral resolving power | N/A | 30 -100 | Higher spectral resolving power at region around 3 μm desirable |
| **Optics** | | | |
| Type of optics | N/A | Al Mirrors, prism as dispersive element ( e.g. Sapphire/MgO) | |
| FOV | Degrees | ±0.5 long<br>5 arcseconds on sky wide | Long slit spectrograph |
| Pixel IFOV | μrad | 10 | |
| Pixel IFOV | m | 10 | At 1000 km range |
| Aperture | mm | 500 | |
| Focal length | mm | 3600 | |
| Focal number | # | 8/4 | Telescope f8, off-axis Korsch , spectrometer and context camera f4 |
| **IR Detector** | | | |
| Type of detectors | N/A | 2048 x 2048 | Teledyne H2RG baselined, European options investigated based on current developments for e.g. Ariel |
| Pixel size | μm | 18 | Operating mode may combine pixels |
| Exposure time | sec | $10 – 8 \times 10^{-3}$ | Dependent on operating mode (point survey/mapping Integrated within instrument during normal operation) |
| Signal to noise ratio | | 50 (goal=100) | Depends on mode (Survey/Flyby) |
| **CCD Detector** | | | |
| Type of detectors | N/A | 2048 x 2048 | European e.g. e2V CCD |
| Pixel size | μm | 18 | Operating mode may combine pixels |
| Exposure time | sec | $10 – 8 \times 10^{-3}$ | Dependent on operating mode (point survey/mapping Integrated within instrument during normal operation) |
| Signal to noise ratio | | 50 | Depends on mode (Survey/Flyby) |
| **Physical** | | | |
| Mass, total | kg | 53 | Based on in f8 telescope and representative CAD model |

| Dimension | mm | Telescope: 520 diameter x 600 long. | Based on f8 telescope and Ariel spectrometer enclosure |
|---|---|---|---|
| Volume | m³ | 0.154 | |
| Operating temperature | K | >120/40 | Optimal telescope temperature via radiative cooling /Detector temperature controlled using local cooler and PID thermostat |
| **Power** | | | |
| Total average power | W | 140 (TBD) | Based on estimates from Ariel instrument study |
| Peak power | W | 140 (TBD) | Based on estimates from Ariel instrument study |
| **Spectrometer Data Volume** | | | |
| Flyby Mode | GBytes | 1.2 | Cumulative total Based on flyby of 10 km asteroid at 1000 km for full spectral maps including space views. No compression assumed. |
| Survey Mode | MBytes/day | 5.0 | Assumes single slit spectroscopy, $R$=100, 40 minute integrations coadded on spacecraft |
| **Visible Context Imager Data Volume** | | | |
| Flyby Mode | GBytes | 1 | Estimated from Rosetta/OSIRIS imaging during Lutetia flyby |
| Survey Mode | MBytes/day | 3 | Assuming 20 64x64pix images/target |

## 6.4 Thermal imager

By mapping an asteroid's diurnal thermal response, a thermal (e.g. 6 – 16 μm) multi-spectral imaging instrument will provide key information on its surface composition and the nature of its surface and near sub-surface (its thermal inertia e.g. buried rock versus dust).

### 6.4.1 Design Description

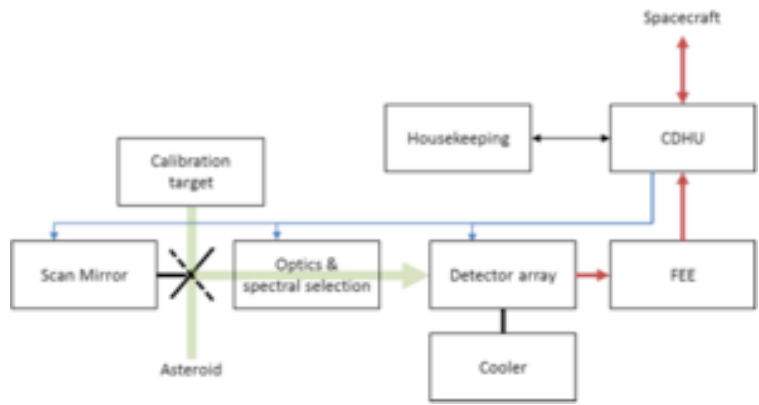

**Figure 15: CMS thermal imager block diagram**

The Thermal Mapper for CASTAway (TMC) is a compact multichannel radiometer and thermal imager based on the Compact Modular Sounder (CMS, Figure 15, Figure 17) instrument currently flying on the UK's TechDemoSat-1 spacecraft in low Earth orbit. The TMC instrument uses a two-dimensional uncooled microbolometer detector array to provide thermal imaging of the asteroid. In addition, fourteen narrow-band infrared filters located around diagnostic mineral spectral features provide additional compositional discrimination (Figure 16). The filters are mounted at an intermediate focus to improve spectral performance. Calibration is maintained by an internal blackbody target and have access to a space view using a single scan/calibration mechanism. Multi-spectral thermal images are generated by push broom scanning of the target during flyby (e.g. Thomas et al., 2014).

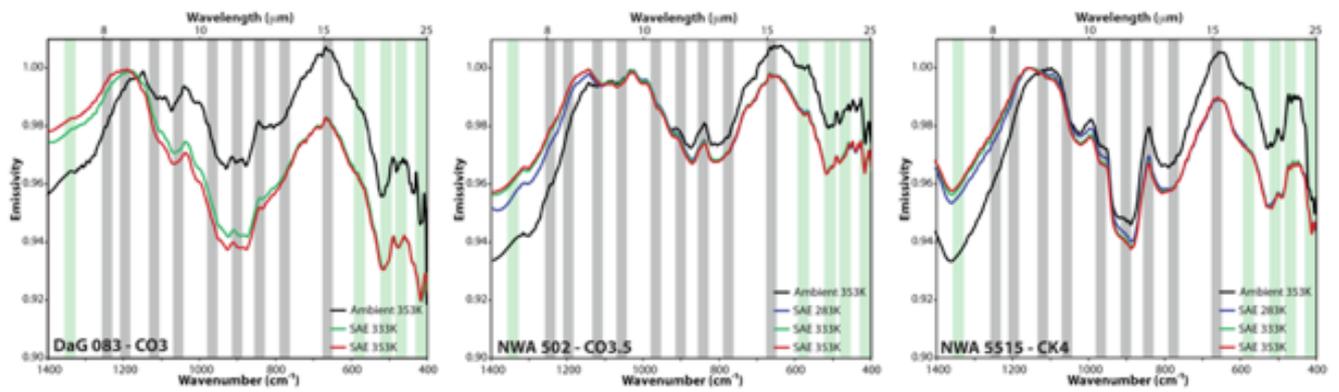

Figure 16: Laboratory emissivity spectra of a range of carbonaceous chondrite meteorites measured under Earth and asteroid-like conditions. Grey and green horizontal bars highlight the band filter locations around key diagnostic feature (after Thomas et al., 2014)

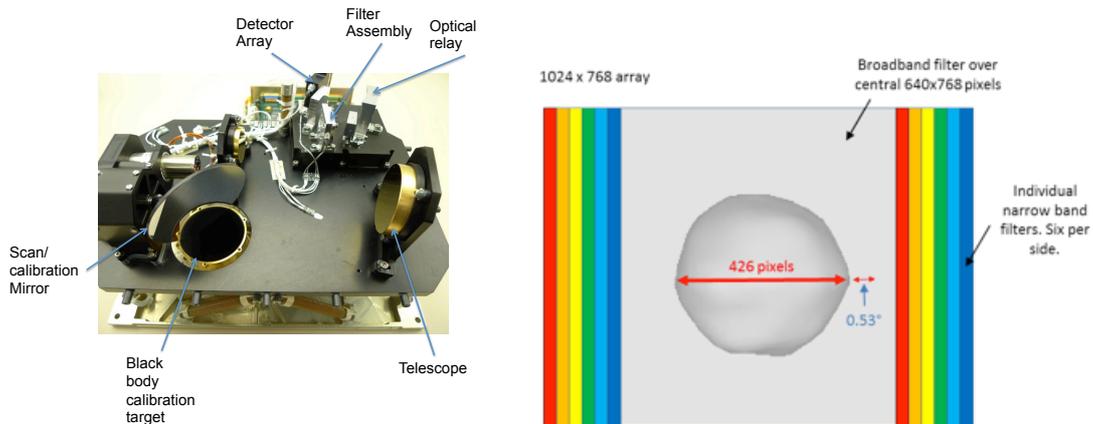

Figure 17. (left ) The TMC heritage instrument, the Compact Modular Sounder (CMS), flight unit minus its outer cover. CMS has approximate dimensions of 380 x 335 x 185 mm. Proposed filter and imaging detector layout (right), example shown during approach phase to the MBC. The TMC will use an optimised layout, reducing the instrument size to 180 x 150 x 120 mm. Mass is 4 kg based on the CMS design.

The heritage instrument, CMS, has dimensions of 380 x 335 x 185 mm Figure 17, left). The TMC radiometer approach has the advantage of a well-calibrated, straightforward data product, and significant flight heritage.

Table 7.  TMC Instrument Summary.

| Parameter | Units | Value/Description | Remarks |
| --- | --- | --- | --- |
| Reference P/L | N/A | Thermal Mapper for Castalia (TMC) | Precision filter imaging radiometer |
| Spectral Range | μm | 5 – 20 | Requires custom detector window |
| Spectral resolution, mineralogy channels | μm | 0.2 | Typical values, based on Diviner flight filters for 8 μm compositional channels |
| Number of Channels | N/A | 11 | 10 narrow band channels, 1 thermal imaging channel |
| Temperature accuracy | K | ±5 (±1 goal) | |
| Temperature range | K | 100 – 200 | Based on Schorghofer (2008) |
| Emissivity accuracy | % | 1 | |
| **Optics** | | | |
| Type of optics | N/A | Aspheric Al Mirrors, multilayer interference filters | |
| Focal number | # | 1.7 | Indication only, the FOV is not circular |
| Aperture diameter | mm | 50 | Based on CMS |

| Detectors | | | |
|---|---|---|---|
| Type of detectors | N/A | Uncooled 1024 x 768 micro bolometer array | ULIS baseline. Other candidate options possible |
| Pixel size | μm | 17 | Operating mode may combine pixels |
| Pixel IFOV | mrad | 0.53 | 17 μm detector, f/# 1.7 |
| **Physical** | | | |
| Mass, total | kg | 4 | Including margin |
| Dimension | mm | 180 x 150 x 120 | Based on consolidation of the CMS mechanical design |
| Volume | cm$^3$ | 283 | |
| Operating temperature | °C | -10 optimal | Can be controlled by a TEC, dependent on detector type used |
| Temperature range | °C | -40 to +60 | |
| Temperature stability | °C | <0.3 per minute | Larger swings can be removed by calibration strategy but not ideal. |
| **Power** | | | |
| Total average power | W | 10 | Based on CMS instrument |
| Peak power | W | 12 | Hot calibration target |
| **Spacecraft Requirements** | | | |
| Pointing requirements | mrad | 1 | Requires space view for calibration |

### 6.5 Asteroid Detection Cameras (ADC)

#### 6.5.1 Measurement Principle/ Detection Concept

CASTAway's baseline Asteroid Detection Cameras are based on an array of μASCs (Micro Advanced Stellar Compasses, Figure 18, Figure 19) to autonomously detect metre-sized objects in the main asteroid belt. Such objects cannot be detected from the Earth and CASTAway will provide us with the first observations of a population of more than $10^{12}$ objects that are a likely source population for meteorites on Earth and generate impacts on other planetary surfaces.

#### 6.5.2 Design description and operating Principle

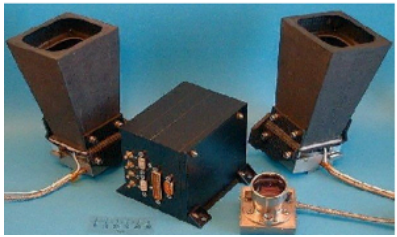

Figure 18: Second generation Advanced Stellar Compass (Credit:DTU)

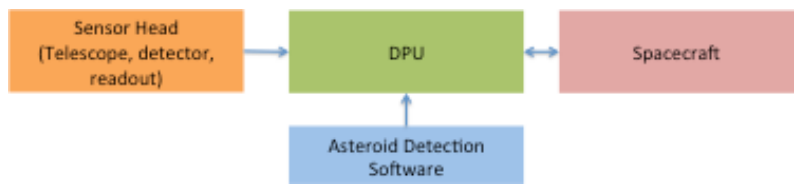

Figure 19: Asteroid Detection Block Diagram

The μASC operates in the following fashion. A star image is acquired at a user specified rate. The image is then sifted for luminous objects (star centroids). If an approximate attitude is known a priori, the centroids measured are fitted to the expected centroids found from the star catalogue. If a good match is acquired, the match process will result in an attitude with an accuracy in the range of 1 arcsecond. If the match procedure fails, or if no a-priori attitude knowledge exists, the μASC will initiate the "lost in space" algorithm, that search the measured centroid list for triplets, i.e. local constellations, that are then matched to a pre-compiled star triplet database. This search takes approximately 70 msec. Based on the solution found, the match procedure to the star catalogue is then repeated. This approach results in a very fast and extremely robust attitude determination.

The measured attitude thus measured may then undergo some user-selected transformations. If desired, the µASC may remove the effect from astronomical aberration, an effect that otherwise will result in a bias of 20 – 30" relative to the J2000.0 frame. Also, the reference frame for the attitude may be transformed into spacecraft coordinates rather than camera coordinates. The field of view for the ADC for CASTAway is set so as to be sensitive to objects down to V = 16.

#### 6.5.2.1 Autonomous detection of non-stellar objects

Because the µASC analyses all luminous objects in the FOV and matches these to the catalogue, it will also detect which objects are NOT stars, i.e. non-stellar objects. At V = 7, virtually all objects thus detected are satellites, planets and planetesimals from previous ground based observations with the µASC. An asteroid such as Vesta is thus reliably tracked throughout its passage of the FOV. It is worth noting, that this process gives the object position directly in right ascension and declination. At V = 9, the list of non-stellar objects from each image include several galaxies, nebulae etc. as well.

### 6.5.3 Asteroid detection with CASTAway

The CASTAway spacecraft will be equipped with 4 star trackers that cover four adjacent sectors around the boresight of the main telescope (Figure 11). While the main telescope observes an asteroid, the star trackers will search for small objects in the same direction as the targeted asteroid. Each observation by the main telescope typically can take from 20 – 60 minutes to several hours depending on brightness. After the observation the spacecraft will re-orient to observe a new asteroid on the prioritized list on the main computer.

A list of moving objects, detected within the four sectors by the µASCs, will be sent to the main computer. Based on pre-selected criteria, some of the detected objects could trigger target-of-opportunity procedures to be observed by the main telescope.

### 6.6 Radio Tracking Science

The Radio Science experiment will make use of the basic TTC subsystem of the spacecraft. An asteroid's mass can be determined to high precision by analysing the spacecraft's radio tracking data (distance/range and velocity/range rate or Doppler frequency). The gravitational attraction of the asteroid acting on the spacecraft during the flyby steadily perturbs the flyby trajectory by an amount proportional to the mass of the asteroid. The magnitude of the perturbation also depends on the flyby distance, the relative velocity between spacecraft and asteroid, and the geometry (Paetzold et al., 2001). A slow velocity is favourable and the angle between the line of sight (LOS) and the direction of the spacecraft's velocity should be between 60° and 120°.

During the flyby the main telescope will point towards the asteroid, which is in conflict with the pointing requirements of the Radio Science experiment. This will result in a loss of tracking data during the closest approach. Since the closest approach contains the highest change in velocity the loss would have a significant impact on the quality of the radio science measurements (Paetzold et al., 2010). Therefore, a steerable antenna shall be used in order to make it possible that the antenna points to the Earth while the main telescope is able to observe the asteroid at the same time. The required pointing precision of the antenna shall ensure that the power loss due to mispointing is less than 0.1 dB.

The scientific goal is the determination of the mass of the asteroid and consequently its bulk density. Therefore, a volume estimate from imaging will be required with accuracy in the same order of magnitude as the mass determination accuracy; this has been demonstrated in previous flyby missions (e.g. Rosetta at Lutetia - Paetzold et al., 2010).

### 7. Survey performance

To assess how well the proposed mission will address the science goals, we use the best available estimate of the total number of asteroids (Bottke, et al., 2005). Statistical distributions (Table 8) are calculated using this model (logarithmic bins, e.g. '10m' includes ~3-30m diameter asteroids) and assuming an even distribution from 2.1 – 3.2 AU, 2 AU thick, an albedo of 10% for all objects, and no phase angle correction. At V=13, there is a target available in a given size range somewhere on sky at any time, although 1-10m objects will not be known and can only be targets of opportunity discovered by the ADCs. Maximum distances are set by magnitude detection limits (V=15 for MTC, V=16 for ADCs), which, along with the FOV, define a volume for detection for the ADCs. The number of possible MTC targets includes a Sun exclusion angle of 60°, and folds in the expected completion (number known / total) at each size bin following 10 years of LSST.

Table 8. Asteroids seen based on statistical model, using FOV and mag. limits of MTC and ADCs. Sizes and distances in km

| Diameter (km) | 1E-03 | 1E-02 | 1E-01 | 1E+00 | 1E+01 | 1E+02 |
|---|---|---|---|---|---|---|
| # in MAB / bin | 2.7E+14 | 6.7E+11 | 2.0E+09 | 1.3E+07 | 1.4E+05 | 1.8E+03 |
| Distance to nearest | 4.8E+03 | 3.5E+04 | 2.5E+05 | 1.3E+06 | 6.0E+06 | 2.5E+07 |
| V mag. of nearest | 12.7 | 12.0 | 11.2 | 9.9 | 8.2 | 6.3 |
| Max MTC distance | 1.4E+04 | 1.4E+05 | 1.4E+06 | 1.4E+07 | 7.5E+07 | 7.5E+07 |
| # observable / time | 18 | 46 | 135 | 885 | 1495 | 20 |
| # known obs. / time | 0 | 1 | 61 | 819 | 1495 | 20 |
| Max ADC distance | 2.2E+04 | 2.2E+05 | 2.2E+06 | 2.2E+07 | 7.5E+07 | 7.5E+07 |
| # in ADC FOV / time | 2.0E-02 | 4.9E-02 | 1.5E-01 | 9.6E-01 | 4.1E-01 | 5.3E-03 |
| # found in mission | 2.1E+03 | 1.8E+03 | 4.4E+03 | 2.8E+04 | 1.2E+04 | 1.6E+02 |

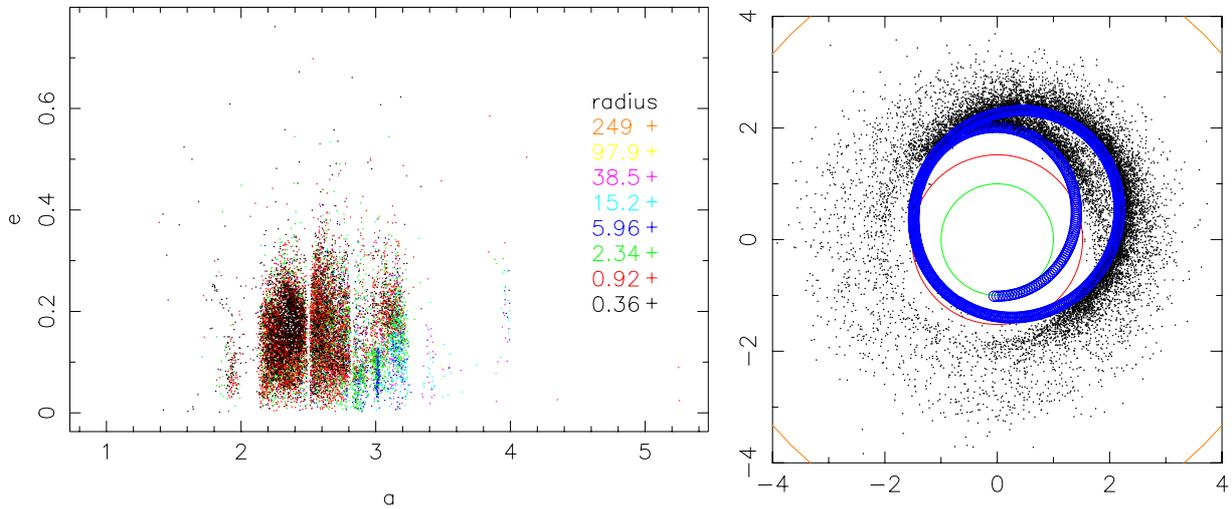

Figure 20. Left: Orbital elements and sizes for all observed asteroids in survey. Right: Position at the time of observation, with Earth, Mars and Jupiter orbits, and CASTAway trajectory (blue).

To look at a 'real' survey, we took the position of spacecraft with time from the baseline trajectory and generated relative positions of all currently known asteroids (~730 000), and from that the apparent magnitudes (including H-G type phase function). Asteroids were classed as observable when brighter than V=15 and at solar elongation > 60°. The survey skips the days within +/- 7 days of a flyby, but otherwise assumes one observation per hour, for 80% of time (i.e. includes conservative margin for calibration, spacecraft activities etc.). The survey logic is to observe the faintest objects that are observable in each period that have not already been seen by the survey (i.e. no repeat observations in this simple set up).
The total number of asteroids seen in this survey is 18 709, including nearly all of the larger ones, which can be seen from relatively large distance, but dominated by the more common sub-km ones seen from < 0.1 AU. The full MAB is sampled (Figure 20), although we see fewer small ones at large semi-major axis and small eccentricity, as they do not come close (N.B. there are also fewer of this category currently known).
It is worth noting that the expected order of magnitude increase in the number of known asteroids from Gaia and LSST before CASTAway launch will mean there are many more targets to pick from in the real survey, especially smaller ones, so it is expected that the 'quiet' times for the survey will be filled in. The maximum number of asteroids that could be observed over the mission with this observing rate is ~44 000.

## 8. Conclusions

We have described a mission that can provide a comprehensive inventory and survey of the Main Asteroid Belt with a scientific payload capable of providing remote sensing of many thousands of point sources to spatially resolved objects during flybys. By utilising an optimised trajectory that includes a significant period of time in the MAB the CASTAway mission concept combines a survey of a large number (several thousands) of point source objects with spatially resolved observations during multiple fast flybys of at least ten. The mission concept was the subject of a detailed study with engineers at OHB System AG, Bremen and it was found that a suitable spacecraft could be built, launched and operated within the constraints of a typical medium sized mission call using either existing launch vehicle capabilities (e.g. Soyuz) or likely future options (e.g. Ariane 6.2). The mission payload suite comprises a survey telescope and spectrometer (0.6 to 5 μm, $R = 30 - 100$) for both close flyby hyperspectral mapping and point spectroscopy during cruise between targets, a multispectral visible context imager, multispectral thermal imager and asteroid detection cameras. Combined with its unique trajectory this payload will allow CASTAway to map variations in composition and size distribution across the Main Asteroid Belt in fine detail for the first time, helping to constrain models of Solar System evolution.

## 9. Acknowledgements

We would like to thank the following organisations for their support in preparing this study:

J. Arnold and N Bowles were supported by Leverhulme Trust grant RPG-2012-814, P. Sanchez was supported by the UK Space Agency for support in developing the trajectory analysis tool kit (NSTP2-GEI1516-020), C. Snodgrass was supported by an STFC Ernest Rutherford fellowship.

The system and subsystem support of the following key persons at OHB System AG during the week long concurrent engineering design session for CASTAway: Bastian Burmann, Ingo Gerth, Alessandro Grasso, Katarina Laydan, Gerrit Proffe, Jan-Christian Meyer, Arne Winterboer and Rolf Janovsky. We would also like to thank Winfried Posselt for their assistance and comments on the optical aspects of the payload.

STFC/RAL Space for their assistance with the initial telescope optical design, cooler designs and telescope thermal analysis.

Giacomo Curzi, Marc Tora, Mario Cano and Rita Neves for their contributions to the trajectory design during their studies at Cranfield University, UK.
Massimiliano Vasile, Marilena Di Carlo and Annalisa Riccardi, at the University of Strathclyde, UK, for their interest and support in the trajectory analysis.

All authors would like to acknowledge and thank their respective institutes for their support during all phases of the preparation of this study.

We thank E. Cloutis and an anonymous reviewer for their comments and improvements to the paper.